\documentclass[journal]{IEEEtran}
\usepackage[export]{adjustbox}
\usepackage{amsmath}
\usepackage{amssymb}
\usepackage{pifont}
\usepackage{rotating}
\usepackage{dblfloatfix}
\usepackage{multirow}
\usepackage{xcolor}
\usepackage{graphicx}% subfigure
\usepackage{subcaption}
\usepackage{enumerate}
\usepackage{MnSymbol,wasysym}
\usepackage{colortbl}
\usepackage{hyperref}
\usepackage{forest}
\usepackage[style=ieee, citestyle=numeric-comp]{biblatex}
\usepackage[ruled,vlined,linesnumbered]{algorithm2e}
\usepackage{titlesec}
\renewcommand*{\bibfont}{\small}
\usepackage{enumitem}
\usepackage{tikz,tikz-3dplot}
\usetikzlibrary{3d,decorations.text,shapes.arrows,positioning,fit,backgrounds, shapes, arrows,arrows.meta, intersections}
\usetikzlibrary{decorations.markings, decorations.text}
\usetikzlibrary{shapes.geometric,positioning}
\usepackage{pgfplots}
\usepgfplotslibrary{fillbetween}
\usetikzlibrary{calc}
\usepackage{gnuplottex}
\usepackage{tikz-3dplot}
\usepackage{lscape}
\usepackage[normalem]{ulem}
\usepackage{fancyhdr}
\bibliography{myreferences}
\definecolor{cerisepink}{rgb}{0.93, 0.23, 0.51}
\definecolor{color1}{RGB}{255, 253, 204}
\newcommand\polygon[5][]%
{   \pgfmathsetmacro{\angle}{360/#2}
    \pgfmathsetmacro{\startangle}{0 + \angle/2}
    \pgfmathsetmacro{\y}{cos(\angle/2)}
    \begin{scope}[#1]
        \foreach \i in {#4}
        {   \pgfmathsetmacro{\x}{\startangle + \angle*\i}
          \fill[#5] (0, 0) -- (\x:#3) --  (\x + \angle:#3) -- cycle;
        }
        \foreach \i in {1,2,...,#2}
        {
          \pgfmathsetmacro{\x}{\startangle + \angle*\i}
          \draw (0, 0) --  cycle;
        }
    \end{scope}
}
\makeatletter
\tikzset{circle split part fill/.style args={#1,#2}{%
 alias=tmp@name, % Jake's idea !!
  postaction={%
    insert path={
     \pgfextra{%
     \pgfpointdiff{\pgfpointanchor{\pgf@node@name}{center}}%
                  {\pgfpointanchor{\pgf@node@name}{east}}%
     \pgfmathsetmacro\insiderad{\pgf@x}
      \fill[#1] (\pgf@node@name.base) ([xshift=-\pgflinewidth]\pgf@node@name.east) arc
                          (0:180:\insiderad-\pgflinewidth)--cycle;
      \fill[#2] (\pgf@node@name.base) ([xshift=\pgflinewidth]\pgf@node@name.west)  arc
                           (180:360:\insiderad-\pgflinewidth)--cycle;            %  \end{scope}
         }}}}}
 \makeatother
\makeatletter
\newcommand\Xsubsubsection{\@startsection{subsubsection}{3}{\z@}%
                                     {-3.25ex\@plus -1ex \@minus -.2ex}%
                                     {1.5ex \@plus .2ex}%
                                     {\normalfont\normalsize\leftskip 0ex}}
\renewcommand\subsubsection[1]{\Xsubsubsection{#1}\leftskip 0ex}
\makeatother
\begin{document}

%\include{AuthorsResponseToReviewers}
%-------------------------------------------------------------
% \hyphenation{op-tical net-works semi-conduc-tor}
%-------------------------------------------------------------
%-------------------------------------------------------------
\title{Towards Understanding Human Functional Brain Development with Explainable Artificial Intelligence: Challenges and Perspectives}
%-------------------------------------------------------------
\author{Mehrin Kiani$^{1}$,
      Javier Andreu-Perez$^{1*}$
      Hani Hagras$^1$,
      Silvia Rigato$^1$,
      and Maria Laura Filippetti$^1$ %
\thanks{$^*$ Corresponding author: \texttt{javier.andreu@essex.ac.uk}}
\thanks{$^1$ Centre for Computational Intelligence, University of Essex, UK}}%
%\thanks{$^2$ Centre for Brain Science, Department of Psychology, University of Essex, Colchester, CO4 3SQ, United Kingdom}}%
\date{}
%-------------------------------------------------------------
\maketitle
%-------------------------------------------------------------
%-------------------------------------------------------------
\thispagestyle{fancy}
\begin{abstract}
The last decades have seen significant advancements in non-invasive neuroimaging technologies that have been increasingly adopted to examine human brain development. However, these improvements have not necessarily been followed by more sophisticated data analysis measures that are able to explain the mechanisms underlying functional brain development. For example, the shift from univariate (single area in the brain) to multivariate (multiple areas in brain) analysis paradigms is of significance as it allows investigations into the interactions between different brain regions. However, despite the potential of multivariate analysis to shed light on the interactions between developing brain regions, artificial intelligence (AI) techniques applied render the analysis non-explainable. The purpose of this paper is to understand the extent to which current state-of-the-art AI techniques can inform functional brain development. In addition, a review of which AI techniques are more likely to explain their learning based on the processes of brain development as defined by developmental cognitive neuroscience (DCN) frameworks is also undertaken. This work also proposes that eXplainable AI (XAI) may provide viable methods to investigate functional brain development as hypothesised by DCN frameworks. 
\end{abstract}
%-------------------------------------------------------------
%-------------------------------------------------------------
\begin{IEEEkeywords}
Explainable Artificial Intelligence, Developmental Cognitive Neuroscience, xMVPA, fNIRS, EEG
\end{IEEEkeywords}
%-------------------------------------------------------------
%-------------------------------------------------------------
\section{Introduction} \label{sec:Introduction}
Human brain development is a complex and dynamic process that begins prenatally and extends through to late adolescence \cite{Stiles_2010_BrainDevelopment}. The human brain has an estimated 100 billion neurons at birth \cite{Ackerman_1992_webpage} whose interconnections form neural networks, which become specialised over time and mediate the functional capabilities of the human brain \cite{Johnson_2001}. This specialisation results not only from the structural development of the brain but also as a consequence of optimisation of inter-regional interactions 
%(known as functional connectivity (FC) \cite{Martijn_2010_functionalconnectivity}) 
in the developing brain \cite{Johnson_2001}. Over the past 50 years, the field of developmental cognitive neuroscience (DCN) has examined the relations between the structural and functional development of the human brain \cite{Munakata_2004}, elucidating the developmental mechanisms underlying cognitive processes such as perception, attention, memory, and language.

DCN research can inform us about the influence of genetic variations and environmental factors in the specialisation of neural networks \cite{Johnson_2001}. In addition, DCN studies can extend insights into how these specialised networks mediate newly acquired social and cognitive functions, shedding light on typical and atypical trajectories of human brain development \cite{KarmiloffSmith_1998}. A greater understanding of brain development trajectories can have profound implications for early detection and the subsequent intervention of developmental disorders \cite{Munakata_2004}. Furthermore, a better understanding of the interplay between structural and functional brain development can be leveraged to inform clinical, educational
and social policies \cite{Johnson_1999}.

In order to examine the neural underpinnings of cognitive processes and their changes across development, functional Near-Infrared Spectroscopy (fNIRS) \cite{Wilcox_2015} \cite{LloydFox_2010}, and Electroencephalogram (EEG) \cite{Dereymaeker2017_EEGReview} have been widely used in DCN studies with infants and children. These neuroimaging modalities are both non-invasive, portable, wearable, and relatively inexpensive compared to functional magnetic resonance imaging (fMRI), which has instead proved pivotal in adult brain neuroimaging. In particular, fNIRS and EEG allow for the young participants to stay engaged in tasks whilst recording their brain activity in more naturalistic postures (e.g., sitting upright vs laying down) and ecologically valid settings such as their homes \cite{LloydFox_2010}. Nevertheless, fMRI has been successfully used in developmental studies with asleep infants \cite{Fransson_2009_fMRI_asleepinfants, Blasi_2011}, and more recently with awake infants \cite{Deen_2017_VisualCortex, Ellis_2020_fMRIawakeinfants}. As fNIRS and EEG are considered the most commonly used and `infant-friendly' modalities to investigate neural substrates in DCN studies, the present review paper will focus on these two modalities and their respective data analysis paradigms.

\begin{figure*}[t] % L B R T -- trim arguments
%----------------------------------------------------------
\subfloat[An infant wearing fNIRS cap.]{\label{Fig:fNIRsbaby}\includegraphics[trim={0cm 0cm 0cm 0cm}, clip=true, width=5cm,height=4cm]{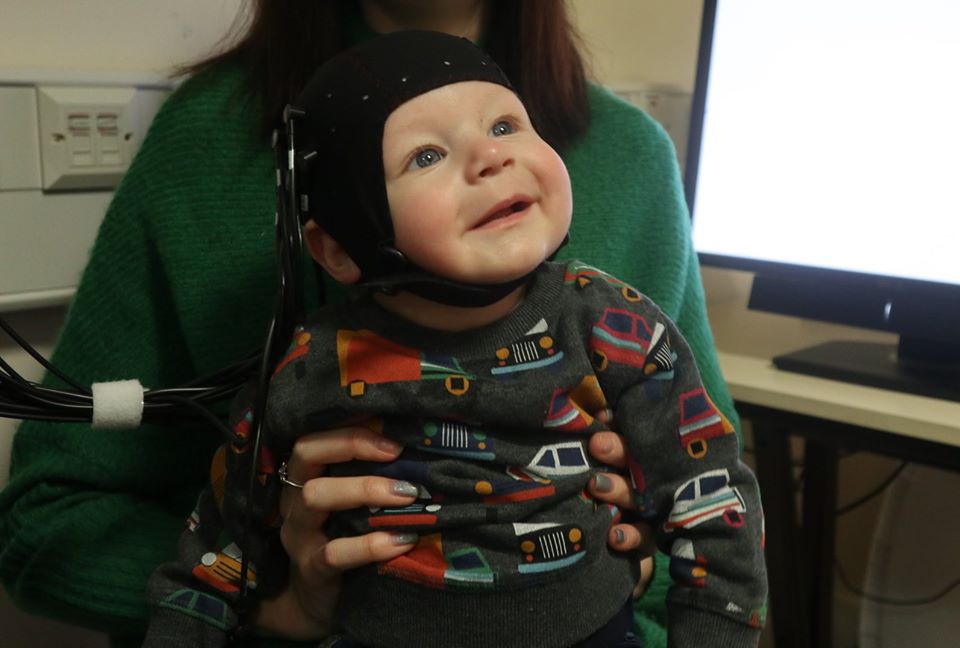}} \hspace{.8cm}
%----------------------------------------------------------
\subfloat[fNIRS principle.]{\label{Fig:fNIRsprin}\includegraphics[trim={7cm 7.4cm 20cm 4cm}, clip=true, scale =.7]{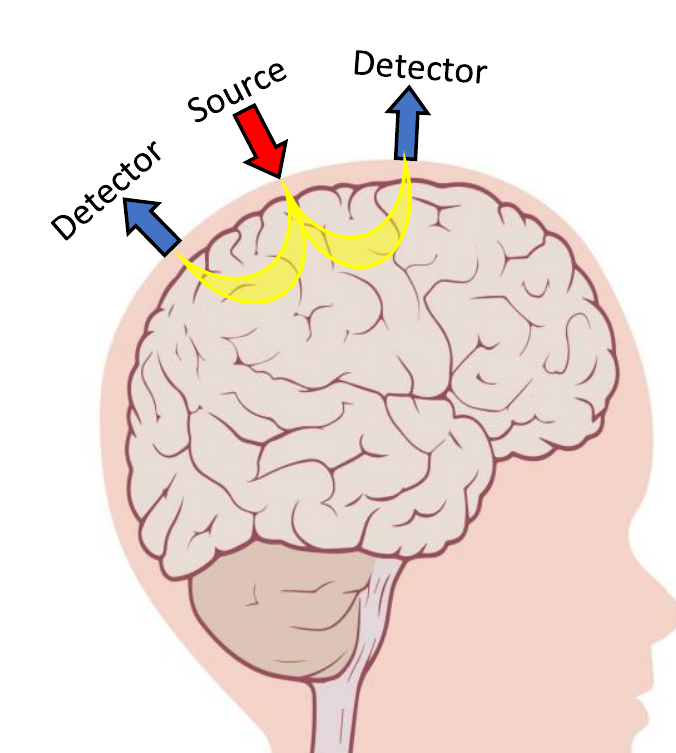}} \hspace{1.9cm}
%----------------------------------------------------------
\subfloat[A fNIRS signal. ]{\label{Fig:fNIRssig}\includegraphics[trim={5cm 10.5cm 3cm 8cm}, clip=true, scale =0.41]{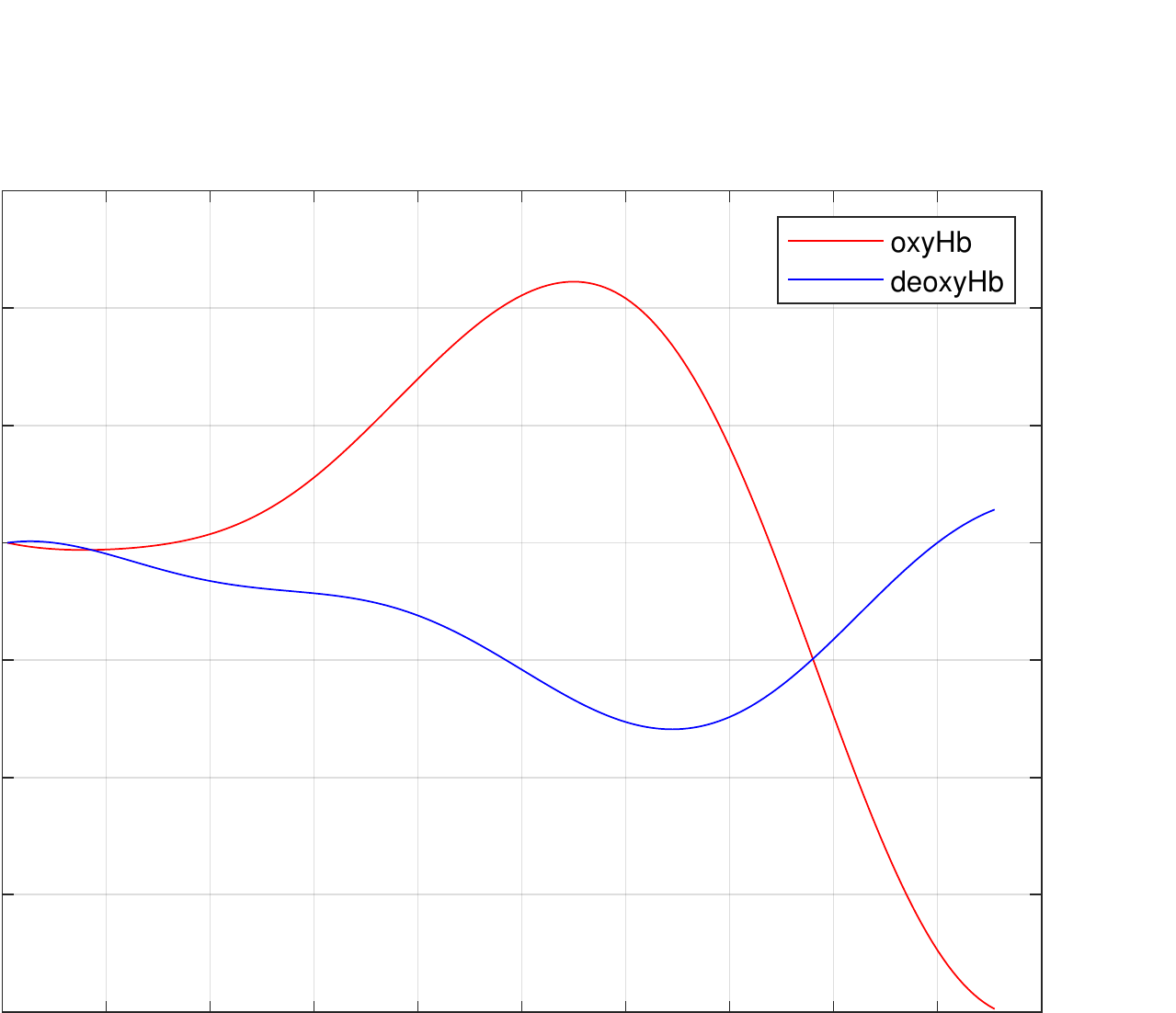}} \hspace{.2cm}

%----------------------------------------------------------
\subfloat[An infant wearing EEG net.] {\label{Fig:EEG1020}\includegraphics[trim={0cm 0cm 0cm 0cm}, clip=true, width=5cm,height=4cm]{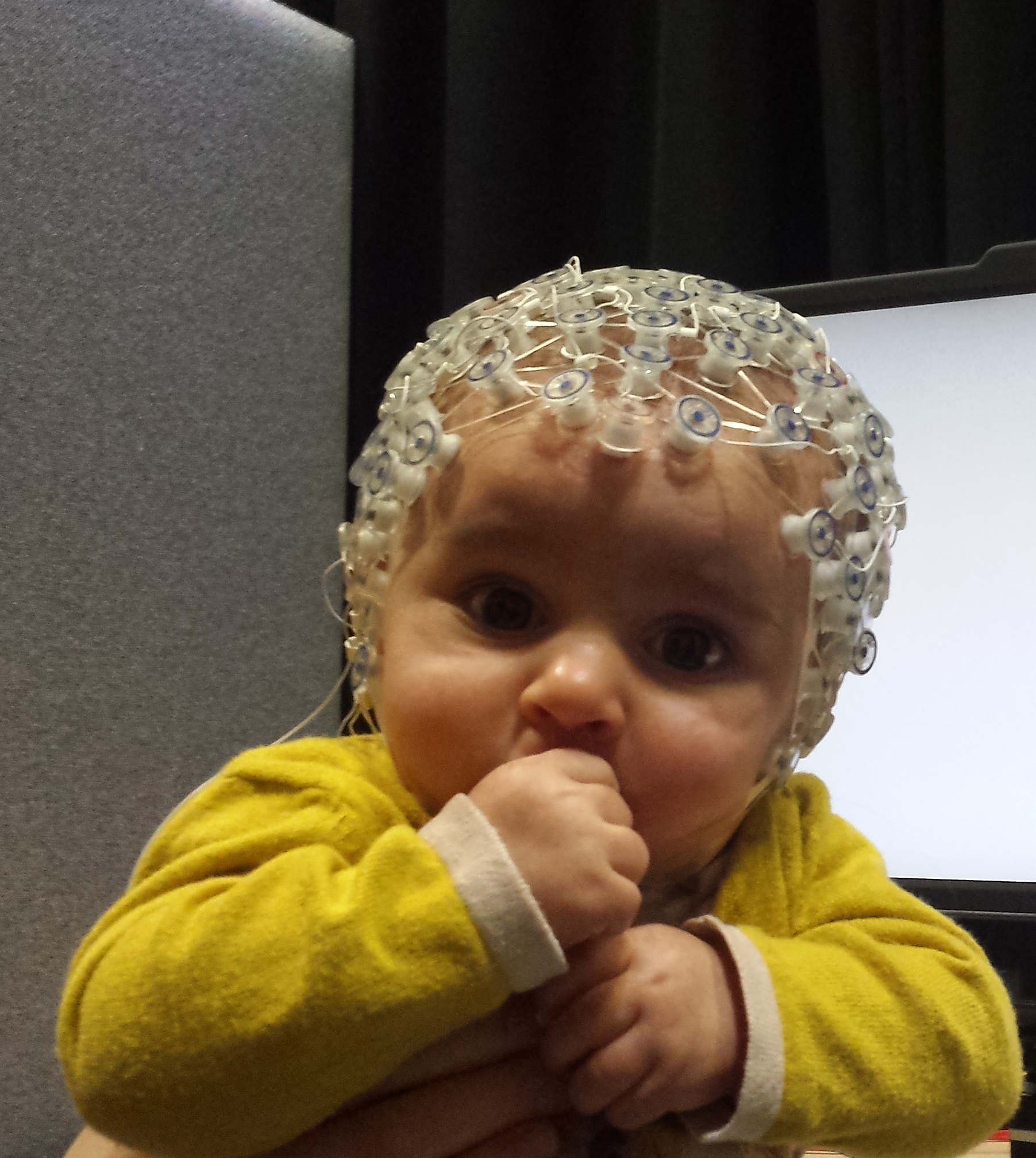}} \hspace{1cm}
% \subfloat[\scriptsize 10/20 EEG electrode system.] {\label{Fig:EEG1020}\includegraphics[trim={17cm 5cm 6cm 4cm}, clip=true, scale =0.45]{Figures/fNIRS_EEG_illustration.pdf}} \hspace{.2cm}
%----------------------------------------------------------
\subfloat[EEG principle.]{\label{Fig:EEGprin}\includegraphics[trim={11cm 7.1cm 15cm 4.5cm}, clip=true, scale=.7]{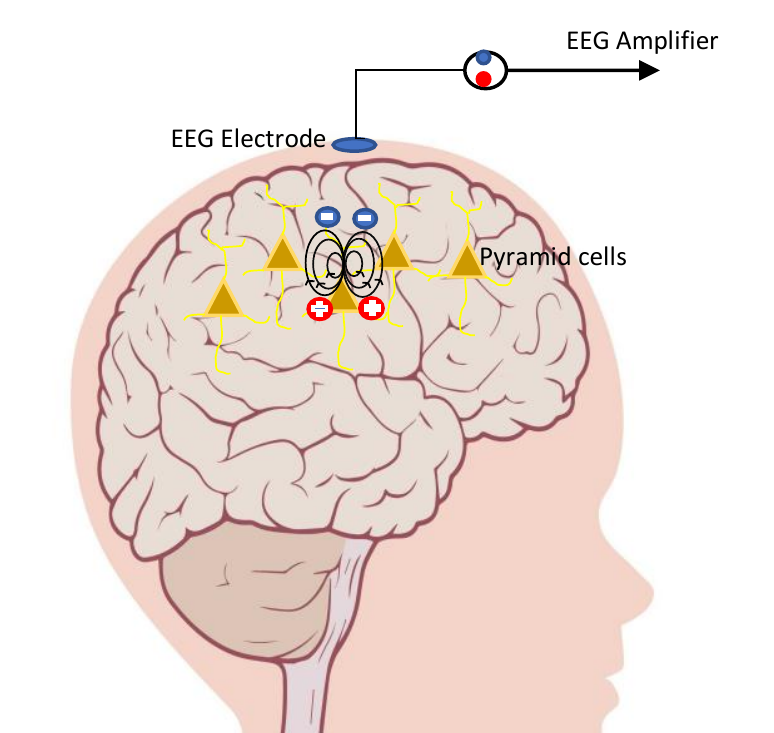}}\hspace{1cm}
%----------------------------------------------------------
\vspace{0.5cm}\subfloat[10 EEG signals.]{\label{Fig:EEGsig}\includegraphics[trim={1cm 1.1cm .1cm 0cm}, clip=true, scale =0.36]{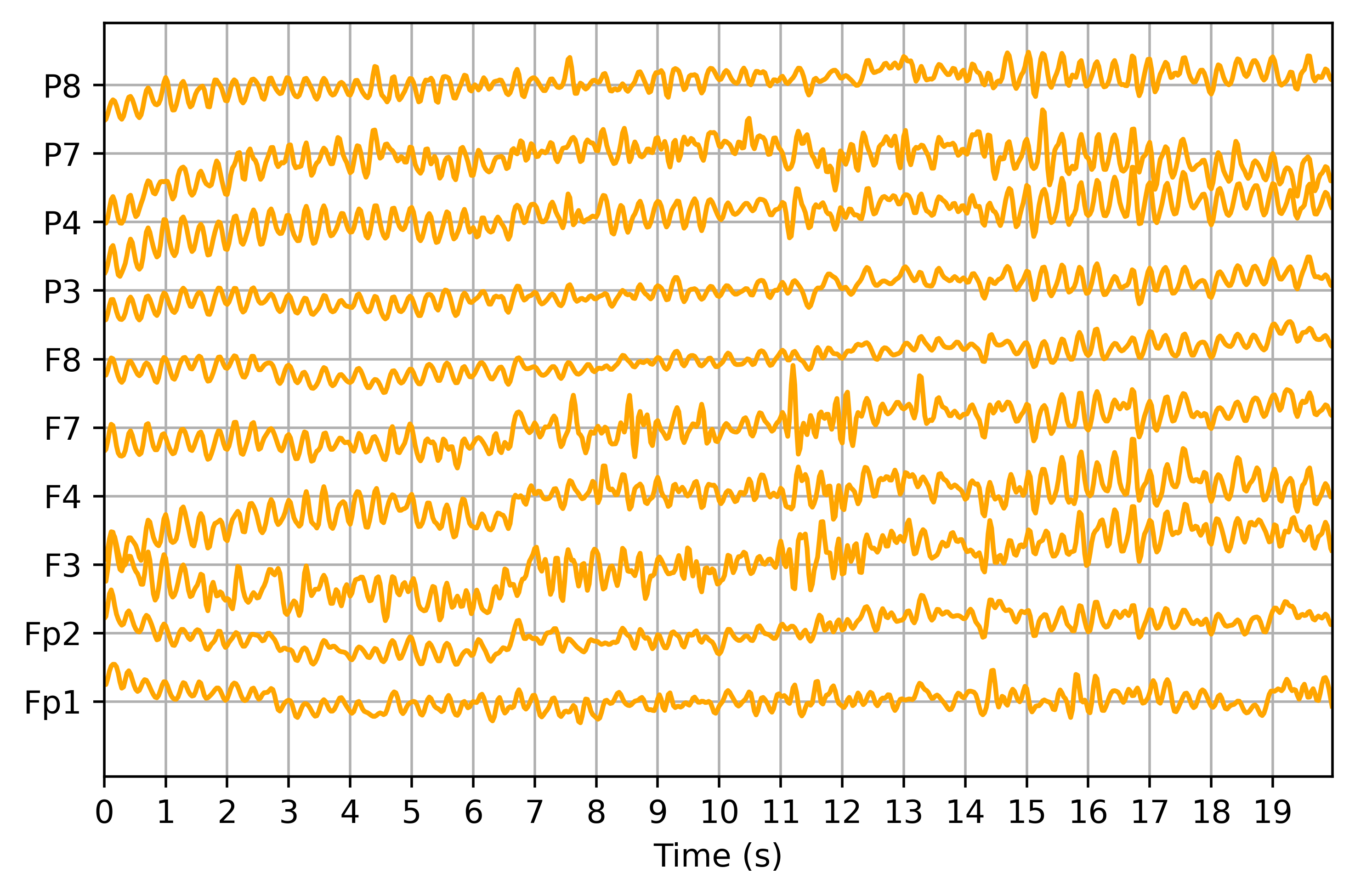}} 
%----------------------------------------------------------
\caption{(a) An infant wearing an fNIRS cap. The placement of the fNIRS channels on the fNIRS cap is dependent on the areas of interest for investigating the underlying brain activity. (b) An illustration of the fNIRS principle: fNIRS is an optical neuroimaging modality that reads underlying cerebral activity using fNIRS channels formed by a pair of sources and detectors. Based on the area under investigation, a fNIRS source shines Near InfraRed (NIR) light at the point on the surface of the head, and the diffusely refracted NIR light is recorded by a fNIRS detector. The relative changes in haemoglobin (Hb) concentration in the blood measured by fNIRS channels is inferred as cortical brain activity \cite{Wilcox_2015}. The delay in the fNIRS haemodynamic response measurement varies significantly depending on the age of the participants and the specific cognitive/motor task undertaken \cite{Roever_2018_Investigation}. (c) A representative fNIRS signal. The y-axis has \(\Delta\) concentration Hb values recorded at time (x-axis) post stimulus presentation. The red signal is \(\Delta\) concentration in oxy-haemoglobin values whereas the blue signal is \(\Delta\) concentration in deoxy-haemoglobin values. (d) An infant wearing an EEG net. The EEG electrode  placement  has  been  standardised  using an international 10–20 system that uses anatomical landmarks on the skull \cite{Homan_1987_EEG1020System}. (e) EEG measures brain electrical activity, with the electrodes placed on the scalp, which reflect the summated postsynaptic potentials of cortical neurons in response to changing cognitive or perceptual states \cite{Haynes_2006_DecodingMentalStates}. EEG activity is mainly generated by pyramidal neurons in the cerebral cortex that are perpendicular to the brain's surface/electrode on the scalp \cite{Britton_2016}. (f) 10 illustrative EEG signals with measured voltages on the y-axis and time on the x-axis. EEG signals have temporal resolution in milliseconds, providing a near real-time display of ongoing cerebral activity, but with limited spatial resolution due to effects of electric field spread \cite{Nunez_2006_EEGFieldSource}. 
%In addition, EEG is also more prone to motion artifacts a significant impediment with infant neuroimaging.
}
\label{fig:fnirs_eeg_comp}
\end{figure*}

% Meeting notes -- 21 jan
%Say that some works use multimodal as well, benefit of both EEG and fMRI
%NIRS and EEG are becoming more common as well -- section for multimodal imaging? 

%see Fig. \ref{Fig:fNIRs_illustration}
%see Fig. \ref{Fig:EEG_illustration}
%issues: motion artifacts, spatial resolution

fNIRS is an optical neuroimaging modality that uses Near-Infrared (NIR) light on the scalp to record changes in blood haemoglobin that occur as a result of cerebral activity. More specifically, fNIRS measures the relative changes in haemoglobin (Hb) concentration in the blood, based on NIR light absorption by the Hb molecules, which is inferred as a measure of the cortical brain activity \cite{Wilcox_2015}. The fNIRS cap, comprising of pairs of sources and detectors, can be flexibly adapted based on the brain areas of interest (see for an example Fig. \ref{Fig:fNIRsbaby}). The strength of fNIRS lies in its good spatial localisation (within 2cm) that allows for conclusions to be drawn about the localised cortical activity from different anatomical locations of the cortical structures, as recorded by the fNIRS channels located on the participant's head (Fig. \ref{Fig:fNIRsbaby}). An illustration of the fNIRS principle (Fig. \ref{Fig:fNIRsprin}) along with a representative signal (oxy-Hb in red, and deoxy-Hb in blue is shown in Fig. \ref{Fig:fNIRssig}) is shown in Fig. \ref{fig:fnirs_eeg_comp}.

%the localized changes in oxygenated (HbO) and deoxygenated (Hb) cortical hemoglobin that occur because of cerebral activity.10,27 These blood-flow changes, termed the hemodynamic response,28 are the result of coupling between the neurovasculature and metabolic demands of neurons in response to increased activity. f

%----------------------------------------------------------
%----------------------------------------------------------
% fNIRS and EEG illustration figure -- START
%----------------------------------------------------------
%----------------------------------------------------------
%----------------------------------------------------------
%----------------------------------------------------------
%----------------------------------------------------------
% fNIRS and EEG illustration figure -- END
%----------------------------------------------------------
%----------------------------------------------------------

While fNIRS relies on changes in blood oxygenation to measure brain activity, EEG measures electrophysiological brain activation. More specifically, EEG records electrical changes on the scalp\textcolor{blue}{,} allowing the measurement of rapid cognitive processes \cite{deHaan_2003} with high temporal accuracy in the order of milliseconds \cite{Burle_2015_EEGTemporal}. The EEG principle is represented in Fig. \ref{Fig:EEGprin}. The EEG net has electrodes fitted in it using a standardised 10/20 electrode placement system that covers the whole head (see  Fig. \ref{Fig:EEG1020}). Since EEG can record activity on the time scale of underlying neuronal activity, EEG signals (representative EEG signal shown in Fig. \ref{Fig:EEGsig}) are best suited for connectivity analysis.  However, EEG is also more sensitive to motion artifacts \cite{Sweeney_2012} and due to its limited spatial resolution \cite{Nunez_2006_EEGFieldSource}, EEG makes it difficult to map brain electric activity read by the electrodes to their corresponding anatomical regions in the brain.

% Multimodal studies
One of the most recent advancements in neuroscience research is the combined use of neuroimaging techniques (e.g., \cite{Saadati_2019_fNRISEEG}). In particular, in DCN, multimodal imaging can provide a wider picture of functional brain activity by benefiting from the advantages of different neural measures (e.g., EEG-fNIRS \cite{Chen_2015}). Given the complementary characteristics of EEG and fNIRS, i.e., EEG records highly accurate temporal information whilst fNIRS is more spatially localised, multimodal fNIRS-EEG studies enable greater information to be recorded regarding the underlying brain activity. While this represents a step further to better study the developing brain, it does not prove sufficient to translate DCN research to inform typical and atypical trajectories of functional brain development. More specifically, if functional brain trajectories using infant's neuroimaging data can be established with the help of explainable Artificial Intelligence (XAI) methods, it may assist in early identification of, and thus intervention in, neurodevelopmental disorders \cite{French_2018_SysRev}, as well as shape policies in the context of typical neurocognitive development.

%Ganea_2018_Development,Schel_2017_Specialization
% Why is it important to know about functional brain development
Indeed, the fundamental question in DCN of how cognitive development is mediated by structural maturation (i.e., the emergence of faculties through growth processes) and optimised interactions remains open. In this regards an understanding of the theoretical frameworks that can explain the bidirectional relation between the structural and functional development of the human brain is critical. Therefore, in Section \ref{sec:IS}, we firstly summarise the key concepts of the DCN frameworks including the `Interactive Specialisation' (IS) theory \cite{Johnson_2001,Ganea_2018_Development,Schel_2017_Specialization} and the neuroconstructivist approach \cite{KarmiloffSmith_1994, KarmiloffSmith_2009}. A review of the current artificial intelligence (AI) algorithms, as applied to fNIRS and EEG data both in infancy and adulthood is undertaken in Section \ref{sec:ML}. The aim of the review is to investigate the extent to which these AI methods can explain human functional brain development in light of the theoretical frameworks of DCN. Implications of explainable AI methods, that also mimics the mechanisms proposed by DCN frameworks, and the conclusion are presented in Section \ref{sec:discussion} and \ref{sec:conclusion} respectively.  

%---------------------------------------------------------------
%---------------------------------------------------------------
\section{Developmental Cognitive Neuroscience (DCN) Frameworks} \label{sec:IS}
%---------------------------------------------------------------
\definecolor{cadmiumred}{rgb}{0.89, 0.0, 0.13}
\definecolor{amber}{rgb}{1.0, 0.75, 0.0}
\begin{figure*}[!t]
\centering
\input{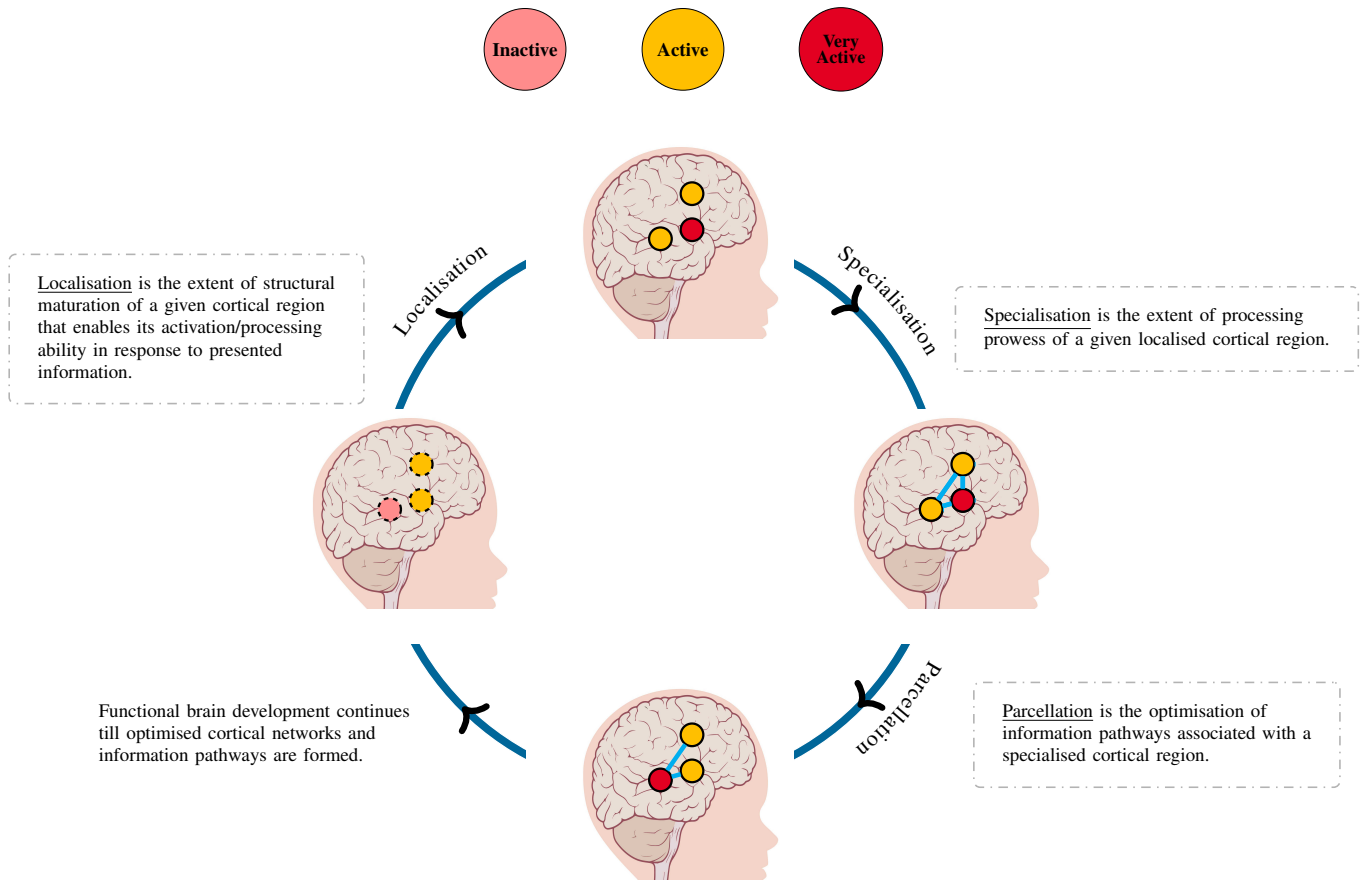}
\caption{The Interactive Specialisation (IS) theory focuses on explaining how different anatomical regions in a developing brain learn to cooperate to form an optimised cortical network using \emph{localisation}, \emph{specialisation}, and \emph{parcellation} processes. The IS theory hypothesises that functional brain development is not a stationary, one-way process, but instead, all of its components are in a loop, giving feedback to each other as structural maturation, as well as external stimuli, pave the way for an increasingly specialised cortical network. An illustration of hypothetical developing cortical activation and interactions are shown in cyan with the circles representing different cortical regions, with varying levels of activity as denoted by their colour. Red: Very active; Amber: Active; and Pink: Inactive. The different levels of activity of a cortical region represent the amount of neurons firing and forming synaptic contacts with other neurons to process presented information. }
\label{Fig:FBD_IS}
\end{figure*}
%-------------------------------------------------------
%-------------------------------------------------------

% Review the section based on: When Younger Learners Can Be Better (or at Least More Open-Minded) Than Older Ones  10.1177/0963721414556653

%https://www.pnas.org/content/114/30/7892

%https://arxiv.org/pdf/2005.02880.pdf <-- read it to see how they are defining the terms

The developed adult human brain, both in terms of structure and function, is a `small world' network \cite{Bullmore_2009}. A small world network is typically characterised with concentrated local activity, decreased short-range interconnections (segregation), and increased long-range connections (integration) rendering it cost efficient. Repeated processing of certain types of input leads to certain brain networks becoming increasingly proficient and fine-tuned to process that specific information \cite{KarmiloffSmith_2009}. In particular, developmental change in the varying levels of activity across different cortical regions leads to gradual specialisation and localisation observed in the developed human brain \cite{KarmiloffSmith_2009}, as illustrated in Fig. \ref{Fig:FBD_IS}. 
%Although the developing brain is more interconnected, the `spurious connections' in the mature brain are a result of pruning due to the increasing specialisation that occurs with experience and epigenetics \cite{KarmiloffSmith_2011_Modularaization}. 

A developed brain is also modular with respect to functional organisation, i.e., it has a hierarchical network that has the ability to feed processed information from one layer (module) to another. The hypothesis of a more modular developed brain is based on the evidence of top-down and bottom-up information flow. For example, during visual processing, the information in the adult brain flows from the primary area of visual processing (such as occipital cortex) to higher hierarchical levels (such as pre-frontal cortex (PFC)) where the information processed by lower hierarchical levels is integrated \cite{Pendl_2017_HierarchicalBrain, Barrett_2012_EvolutionHumanBrain}. 
%\cite{Sigman_2005_TopDown}. 

%In contrast, DCN works investigating object processing have found infants to struggle with object processing tasks, for example see  \cite{Wilcox_2015_ObjectProcessing_infants}, and suggest that an inadequate top-down feedback pathways might therefore be a key mechanism for processing information in brain.

The three main DCN frameworks, namely 1) Maturational perspective, 2) Interactive Specialisation (IS), and 3) Skill learning, aim to answer the question of how these optimised, hierarchical networks emerge during postnatal development. For the purpose of this work, we will focus on the IS perspective, which is largely supported by DCN studies \cite{Johnson_2010_Book_DCN}. The IS framework proposes that both feed-forward and feedback connections between different cortical regions affect the functional specialisation of cortical regions \cite{Johnson_2010_Book_DCN_ch12}. More specifically, the IS theory provides a description of the following three major processes that occur in the developing brain:
\begin{enumerate}[label=(\roman*)]
    \item \emph{Localisation}: The extent of cortex activation for a given task.
    \item \emph{Specialisation}: The extent of functionality achieved by a given cortical area. 
    \item \emph{Parcellation}: The optimisation of synaptic connections of neural circuits.
\end{enumerate}

The IS framework suggests that functional brain development is a dynamic process with localisation, specialisation, and parcellation processes forming a continuous loop of development as shown in Fig. \ref{Fig:FBD_IS}. As a given cortical area gains more structural maturation, its specialisation for a given task increases, which then triggers the parcellation (optimisation) of information flow in the cortical network formed to subserve that given task. 
%The updated neural circuit formed gives feedback to the cortical area which then updates its activation levels and extent accordingly paving way for more specialised functionality of the cortical area, and so forth till an optimised small world network is formed \cite{Johnson_2010_Book_DCN_ch12}.  
%At the optimisation stage a greater informational isolation of its neural circuits, i.e. parcellation, is achieved. 

Optimisation can take place because of structural and/or functional maturation (i.e., the emergence of capabilities through growth processes) of different parts of the brain, along with more long range connections coming `on line'. As a result of the parcellation process, not all parts of a given cortical region need to be activated nor are all connections required to transmit the information to the next level of processing. In this sense, parcellation takes place both within and between cortical regions. The increased segregation of information pathways gives rise to increased specialisation (i.e., a modular structure), thus leading to the gradual emergence of hierarchical networks.

%-------------------------------------------------
%--------------- The Neural Reuse ---------------
%-------------------------------------------------
\begin{figure*}[!h]
    \centering
    \input{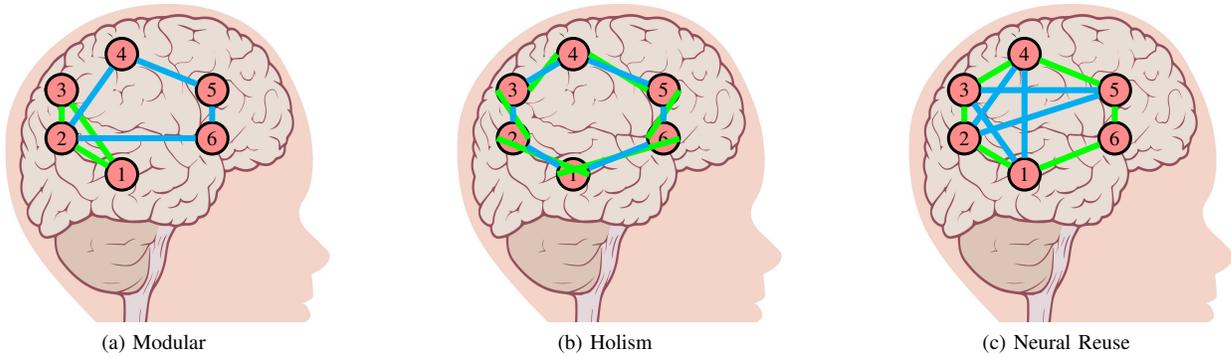}
    \caption{Three different perspectives for the functional structure of the brain are illustrated with cortical modules (denoted by circles numbered 1 to 6) and interconnections between the cortical modules (shown in green for task Y and cyan for task Z). (a) The modular structure suggests that different modules will be activated for different tasks with limited overlap of modules. This is shown by largely segregated modules activated for task Y (i.e., the modules 1, 2, and 3 activated for task Y) and task Z (i.e., the modules 2, 4, 5, and 6 activated for task Z). As such the modules engaged are largely distinct with limited overlap (of module 2 in this hypothetical case) for the two separate tasks Y and Z i.e., brain functional organisation is local. (b) In holistic organisation, all local brain regions are presumed to be involved for all tasks. Hence, all modules are activated for both task Y and task Z. (c) According to the neural reuse perspective, the brain works as a whole but functional differences can be appreciated owing to their reconfigured interactions between the same modules (use and reuse).}
    \label{fig:Neural_Reuse}
\end{figure*}
%-------------------------------------------------
%--------------- The Neural Reuse ---------------
%-------------------------------------------------

% Neural Reuse and Temporal Information
An important consideration with regard to the hierarchical brain is that the interactions between hierarchies at multiple levels and timescales are not hard-wired, i.e., the coordination between modules is not fixed \cite{DSouza_2016_Devpers}. As a consequence, existing modules could subserve emerging cognitive states through a reconfiguration of the evolved circuits using \emph{neural reuse} \cite{Anderson_2010_NeuralReuse} process of brain organisation. The other two plausible processes put forward to explain functional brain organisation are \emph{modularity} and \emph{holism} \cite{Anderson_2010_NeuralReuse}. The modular functional brain structure would imply that for each task there would be largely segregated cortical circuits with limited overlap, whereas, the holism organisation of the brain suggests that all cortical circuits may be engaged across all tasks. The idea of neural reuse seems plausible with respect to optimal usage of existent circuits evolved for a given cognitive task.  In this way, while neural circuits are modular to some extent with respect to their individual functionality, neural reuse suggests that they (individual modules) have the capacity to connect with each other in numerous configurations to achieve a range of cognitive-behavioural tasks. The three aforementioned perspectives of functional structure of the brain are illustrated in Fig. \ref{fig:Neural_Reuse}.

Taken together, the IS framework and neural reuse perspective can shed light on functional brain development at a given time point using cross-sectional DCN studies. However, a major component of functional brain development that is still to be accounted for is the associated temporal information, i.e., at what time the developmental changes are happening \cite{DSouza_2016_Devpers}. Clearly, all stages of functional brain development are not the same with respect to time. In this regard, investigating the temporal dimension associated with a developing brain analysis of longitudinal DCN data, i.e., neuroimaging data recorded over a certain time period, is considered imperative.

%For example learning a new language is relatively easier for a child than an adult. In this regards, the concept of \emph{perceptual narrowing} also plays a significant role in functional brain development since it can account for how early formation of some modules constrains the emergence of future ones. Indeed, perceptual narrowing is a consequence of neural commitment arising from existent modules that defines the formation of new modules. To investigate both temporal information, and perceptual narrowing of a developing brain an analysis of longitudinal DCN studies, i.e. neuroimaging data recorded over a certain time period, is required. 

%
Any AI method used to shed light on functional brain development must keep in mind the aforementioned phenomena and challenges associated with DCN. To this end, %in order to bridge the gap between the analysis of infant's neuroimaging data and theoretical DCN frameworks,
we will be reviewing the extent AI methods can explain their underlying mechanisms to shed light on the processes of functional brain development. %Aside from reviewing the explainability of AI methods to inform functional brain development, the underlying mechanisms of the AI methods will also be discussed from an DCN framework's perspective.

% the work of Alexander Luhmann
%https://scholar.google.com/citations?hl=en&user=SOvqoAgAAAAJ&view_op=list_works&sortby=pubdate
%Using the General Linear Model to Improve Performance in fNIRS Single Trial Analysis and Classification: A Perspective
%---------------------------------------------------------------
%---------------------------------------------------------------
\section{Artificial Intelligence (AI) Methods in Cognitive Neuroscience} \label{sec:ML}
%---------------------------------------------------------------
%---------------------------------------------------------------

The generic field of cognitive neuroscience investigates the underlying brain functional mechanism that subserve cognitive processes such as memory, perception, understanding, and reasoning \cite{Churchland_1998_CN}. DCN is a sub-field of cognitive neuroscience that focuses on developmental population (infants and younger children) to investigate how functional brain developmental processes shape the developing brain. In principle, the AI techniques that have been applied to study the cognition states of non-developmental population (such as adults) can also be used to study the developmental population (infants and younger children). This is because all the pre-processing stages of acquired neuroimaging data (fNIRS or EEG) would be similar as well as the AI techniques that can discern the difference in brain activation patterns for adults should also be able to decode the same in infants neuroimaging data analysis as well. As opposed to cognitive neuroscience for adult populations, due to the lack of prior assumptions or cannon models, the application of XAI in DCN helps to bring new light into a science that otherwise, with classical non-explainable or purely statistical models, would be challenging to elucidate.

AI techniques \cite{Theobald_2017_MLforB} have been both inspired by, and used for the study of the learning processes in the human brain. A major component of functional brain development is attributed to unsupervised learning \cite{Zador_2019_CritiquePureLearning} owing  to the massive amounts of unlabeled sensory data infants receive, although supervised and reinforcement learning faculties are also hypothesised to account for some facets of human brain development \cite{Niv_2009_RL}. There is also considerable debate about how much of the functional brain development is a result of postnatal learning, and to what extent is the genome (an organism's complete set of hereditary material) responsible for shaping brain development %by outlining the sensory representations and innate behaviour
\cite{Zador_2019_CritiquePureLearning}.

%\cite{LeCun_2019_CakeAnalogy}
Considering the aforementioned learning mechanisms in AI methods, and how they can potentially shed light on the human functional brain development, in the following subsections we review the most commonly used AI methods as applied to fNIRS and EEG neuroimaging data. Most studies have not necessarily used these algorithms for the analysis of infants' neuroimaging data, however, their application to infants data would be similar in principle.
%More specifically, upon learning the inference mechanism from a given neuroimaging dataset, the algorithms can inform about the synaptic contacts and increased encapsulation of the cortical networks activated in the brain to perform a given task.
The overarching aim of the review is to investigate the potential and limitations of these algorithms, as applied to infants neuroimaging data analysis, to explain their learnt inference mechanism in terms of developmental brain processes of localisation, specialisation, parcellation, and neural reuse as outlined by the DCN frameworks.

For this reason, here we review AI methods with application(s) to fNIRS and EEG data, as well as some recent promising works for their applicability to DCN research and data analysis. Please note this is not meant to be an exhaustive review of all the AI methods used in cognitive neuroscience studies, nor is it designed to be used as a reference for implementing the reviewed AI methods. The aim of this review is to understand the underlying inference mechanism of the explored AI methods, and what their understating can inform us about the underlying developing brain processes.
 
In this regard, depending on how much can be inferred (or explained by the learnt inference mechanism(s))\textcolor{blue}{,} the AI methods can be generally categorised as explainable (inferred output can be interpreted with linguistic concepts and propositions), partially explainable (inferred output can shed light on the feature importance or rank) or non-explainable (no insight can be obtained from the inferred output). However, with respect to their implications in DCN research, there is not much of a distinction between non-explainable and partially explainable methods\textcolor{blue}{;} hence\textcolor{blue}{,} we will cover the partially explainable methods within the non-explainable methods as well.

The most commonly involved AI methods explainable or not for the analysis of neuroimaging data can also be broadly classified into the following three analysis paradigms: 1) \emph{Connectivity Analysis (CA)}; 2) \emph{Representation Learning (RepL)}; 3) \emph{Multivariate Pattern Analysis (MVPA)}. We briefly summarise these analysis paradigms before reviewing the explainable and non-explainable AI methods as applied to adults' and infants' neuroimaging data. 

%----------------------------------------------------------
\noindent\emph{1) Connectivity Analysis:} 
%----------------------------------------------------------

Brain connectivity analysis can shed light on the segregation and integration of the isolated cortical networks formed to mediate coherent cognitive and behavioral states. 

The three modes of brain connectivity analysis \cite{Fingelkurts_2005_FC} that can inform us about the organisation and the workings of the developing human brain are: 1) structural connectivity (SC) 2) functional connectivity (FC) and 3) effective connectivity (EC) analysis. SC is generally associated with respect to the anatomical wiring in the brain and is typically measured in vivo using diffusion weighted imaging. FC is measured as the temporal correlation between spatially remote neurophysiological events \cite{Friston_2011_FandEC}. In contrast, EC measures the influence that one neural system exerts over another which can be both activity, and/or time dependent \cite{Friston_1993_EC}. 
%The aforementioned modes of the brain activity are illustrated in Fig. \ref{fig:brain_connecitivity} to clear the distinction between them. 

In most cognitive studies, to understand the underlying connectivity of cortical regions for processing presented information, the analysis of FC (to investigate which spatially distinct cortical areas of the brain are engaged simultaneously) and/or EC (to investigate the extent of influence one cortical region exerts on another) is undertaken. Indeed, the analysis of FC and EC can potentially inform about brain architecture; however, to what extent the connectivity analysis effectively contributes to the understanding of brain processes is dependent on the choice of the AI technique (used for the connectivity analysis). %Hence, in the following subsections, the AI methods that have been used to analyse the connectivity data (from fNIRS or EEG) are reviewed to appreciate the extent the learnt analysis mechanism of the AI method can be explained in terms of the analysed connectivity measures.

%----------------------------------------------------------
\noindent\emph{2) Representation Learning:}
%------------------------------------------------

Many recent works in neuroscience are increasingly using deep leaning paradigms to investigate the underlying brain activity in response to a presented task \cite{LeCun_2015_DeepLearning}. Amongst Deep Neural Networks (DNNs), convolutional neural networks (CNNs) have gained particular interest because of their remarkable performance in unsupervised automatic feature extraction and classification of objects in challenging image classification problems \cite{LeCun_2015_DeepLearning}. Owing to the capability of CNNs to compose higher level features using lower level features, CNNs can learn representations of input data automatically overcoming the long standing challenge to handcraft a feature set in conventional AI methods \cite{LeCun_2015_DeepLearning}. In CNNs, a small matrix of numbers (called a filter) is passed over (convoluted with) the raw data, to extract features from the raw data, such as edges in images, also called a feature map. The convolution layer is followed by a pooling layer which downsamples the input to reduce both the spatial size of the input data and the number of hyperparameters in the network. A typical CNN architecture consists of the following stages:
\begin{enumerate}[label=(\roman*)]
    \item Feature Learning Blocks
        \begin{itemize}
            \item Convolution (C) + Rectified Linear Unit (ReLU).
            \item Pooling (P).
        \end{itemize}
    \item Classification/Regression Blocks
        \begin{itemize}
            \item Fully Connected Layers.
            \item Softmax, Logistic regression layer, regression loss (Root Mean Square Error (RMSE) etc.)
        \end{itemize}
\end{enumerate}

The performance of CNNs is critically dependent on the optimisation of hyperparameters, and owing to the large number of hyperparameters that need optimisation, most DNNs, including CNNs, require large datasets to converge. The hyperparameters of a CNN include the size of the filter(s), stride, number of hidden layers, and the learning rate.

%----------------------------------------------------------
\noindent\emph{3) Multivariate Pattern Analysis:}
%------------------------------------------------

In most multivariate analysis, the feature set is crafted by hand i.e., the statistic characteristic (such as the mean or amplitude) of a neuroimaging signal which would best capture the neural underpinnings, corresponding to the task at hand, is chosen manually. The two dimensional matrix formed by collating together the features from \(\mathbf{N}\) channels (for fNIRS) or electrodes (for EEG) and \(\mathbf{J}\) number of data trials is then given as input to an AI method, and is hereby referred to as a multivariate matrix (MVM).

Although it requires considerable subject-matter expertise to curate a feature set for MVM that best represents the underlying neural activity, the classification results based on the analysis of MVM would reflect on the representational dynamics of the underlying cortical networks (as read from fNIRS channels or EEG electrodes). In this regard the classification results obtained from the analysis of MVM can be at least partially attributed to the cortical networks activation as represented by the statistical feature used for constructing the MVM.  

The MVM can be readily analysed using any state-of-the-art AI methods. Most AI methods such as Support Vector Machine (SVM) and Random Forest (RF) usually give very robust classification results with MVM. This analysis approach is termed multivariate pattern analysis (MVPA) \cite{haxby2014} and was first used for neuroimaging analysis on adult multi-voxel fMRI data \cite{Norman_2006_MVPA_fMRI}.

% In particular, the classical analysis paradigm for infants neuroimaging data has been univariate \cite{Aslin_2015_HumanCorrelates} i.e. investigate how brain activity vary at different brain regions, but consider each individual region separately. Although this has immensely increased our knowledge of functions of the different regions of the brain individually, it can not shed light into the interactions of different regions to process presented information. In addition, many 

%add here a introduction to the next sections
In the following subsections, we review the explainable and non-explainable AI methods used on the aforementioned analysis paradigms on both non-developmental (adults) and developmental (infants) population.

\subsection{AI in Cognitive Neuroscience for Adult Brains}
In Cognitive Neuroscience, AI methods are frequently used with adult populations (mature brains). Some approaches can provide no explanation or simply partial information, and others can derive some explainable structure. 
%In the next subsections, we discuss the AI methods, non-explainable and explainable, as applied to adult populations. %
%----------------------------------------------------------
\subsubsection{Non-Explainable AI Methods} \label{sec:NonXAI_Adults}
%----------------------------------------------------------
A review of the non-explainable AI methods for investigating cognitive processes in adults' cognitive neuroscience studies is presented next. 

% \begin{enumerate}[label=(\roman*)]
%     \item FC with EEG using Support Vector Machine (SVM)
%     \item FC with fNIRS using Ridge Regression (RR)
%     \item FC with fNIRS using Interpolated Functinal Manifold (IMF)
%     \item RepL with EEG using EEGNet
%     \item RepL with fNIRS using CNN
%     \item MVPA with EEG using SVM
% \end{enumerate}
%-----------------------------------------------------------
\noindent\emph{a) FC with EEG using SVM}  \label{sec:FC_EEG_SVM}
%-----------------------------------------------------------

The EEG studies by Moezzi \emph{et al.}  \cite{Moezzi_2019_EEG_SVM_FC} and Klados \emph{et al.} \cite{ Klados_2020_PersonalityProfiles_EEG_FC_SVM} used SVM with radial basis function (RBF) as the underlying kernel to investigate EC. In particular, the work by  Moezzi \emph{et al.} is of interest with respect to DCN research as it investigated the difference in FC to recognise young (mean age 24 years) from old adult brains (mean age 71 years). The FC was studied in the standard frequency bands of delta (1–4Hz), theta (4–8Hz), alpha (8–13Hz), beta (13–30Hz) and gamma (30–45Hz). The aim was to study oscillations in standards frequency bands to uncover coordinated activity in large-scale brain networks which facilitate information flow between spatially distributed brain regions. The calculation of FC matrices was done using imaginary coherence in an attempt to account for the poor spatial localisation of the EEG signals.

Cross-validation was performed to optimise the hyperparameters (\texttt{C:} regularisation factor, and gamma kernel coefficient) of SVM, improve accuracy, and identify the most significant features. To map the FC to brain regions, a grouping approach was used to spatially localise the observed connectivity patterns. In addition, consensus features were obtained using Euclidian distance between electrode pairs to investigate FC patterns based on age. They concluded that consensus features belonging to delta, theta, alpha and gamma frequency bands had positive weights showing significantly higher FC in younger adults than in older adults. Features of the beta band had negative weights showing significantly higher functional connectivity in older adults than younger adults. However, as is also acknowledged in the original study \cite{Moezzi_2019_EEG_SVM_FC}, the limitation to map the consensus features to anatomical regions of the brain could not facilitate further discussion on FC pattern differences with respect to brain regions. Hence, despite the prowess of SVM to differentiate between the FC patterns of old and young brains with 93\% classification accuracy, the SVM's inference mechanism could not shed light on the temporal correlations of the different cortical regions.

The non-explainability of SVM inference mechanism is because the learnt support vectors, which form the inference mechanism of SVM for distinguishing between data instances belonging to distinct classes, are defining a hyperplane that optimally separates the data instances in a high dimensional space. Hence, the support vectors as such can not be expressed in terms of the underlying brain activity patterns. The only relevant information that can be obtained from the support-vector of a linear SVM is a feature relevance/significance score but that too can not shed light on the association between the inputs to uncover the cortical networks formed.
%it does not provide either which behaviour of the input is causing their such relevancy, for example because is low and high for each condition etc

%------------------------------------------------
%Regression Model
%------------------------------------------------
\noindent\emph{b) FC with fNIRS using Ridge Regression (RR)} 
%------------------------------------------------

The connectivity analysis with fNIRS does not require additional spatial localisation of the measured cortical activity owing to the relatively good spatial resolution that can be achieved with fNIRS instruments \cite{Wilcox_2015}. Two complementary, non-explainable AI methods, namely ridge regression (RR) and interpolated functional manifold (IMF), used with fNIRS connectivity measures are reviewed next.

A fNIRS study investigating intrinsic FC of cortical networks to predict anxiety states using linear ridge regression (RR) models is done by Duan \emph{et al.} \cite{Duan_2020_FC_fNIRS_CorticalNetworks}. The resting state FC was calculated using Pearson correlation coefficient for 1035 edges between 46 nodes (fNIRS channels). The RR was able to predict the anxiety score with statistical significance using the connectivity of cortical networks. The mean square error (MSE) of their model was 122.04 with correlation coefficient of \texttt{r = 0.36}.

The prowess to predict states of anxiety using FC has profound implications for the diagnosis of anxiety and related disorders. However, the ability for the regression model to explain its 1035 optimal values of \(\beta\) ( also called regressors) in terms of FC is significantly limited. Therefore, despite getting statistically significant results, the FC analysis could not shed light on the resting state cortical networks. 

%----------------------------------------------------
\noindent\emph{c) FC with fNIRS using Interpolated Functional Manifold (IFM)}
%----------------------------------------------------

A recent study that puts forth a solution for group-wise explorative analysis using manifolds is presented by Avila-Sansores \emph{et al.} \cite{AvilaSansores_2020_IFM}. In this work, fNIRS values are projected to an ambient space. Since there can be infinite surfaces that can cross the projected fNIRS values, the aforementioned study proposes Interpolated Functional Manifold (IFM) to select a surface. In particular, an explicit model for the surface is chosen by interpolating between the projected fNIRS values using RBF. 

The proposed IFM method is used on subjects with varying levels of surgical expertise (knot-tying). The fNIRS values are projected onto a two-dimensional manifold and the distribution of the fNIRS values is based on pairwise distances i.e., points that are close together in the manifold have similar characteristics. For this particular study, the medical students' fNIRS responses got projected to the edges of the manifold, whereas more experienced participants' (trainees and consultants) fNIRS responses accumulated in the conceptual centre of the manifold. The graphs were validated against mixed effect models (with regressors encoding group variances) and psychophysiological interaction (PPI). Since IFM analysis may contain infinite graphs, they visualised the FC with IFM graphs by thresholding them to obtain maximum similarity of Jaccard Index(JI). The maximum JI values reported with group level analysis are \(0.89 \pm 0.01\) and those with PPI are \(0.83 \pm 0.07\).

% may be  move this to discussion?:
The advantage of IFM approach is that an explicit analytical expression is obtained that can be used to quantitatively study the group based differences, as in the case of participants with varying level of expertise for certain motor skill. In addition, the IFM approach can facilitate fNIRS data analysis in hyper-scanning studies, i.e. blue, reading neuroimaging data from more than one person at a given time. However, it is a complimentary analysis for measuring FC since the graph of FC measures is selected by thresholding it against established group level models to obtain maximum values of JI.

%--------------------------------------------------------
\noindent\emph{d) RepL with EEG using EEGNet}
%--------------------------------------------------------

In this section, we review the works that learn representations of input data with multiple levels of abstraction, using CNNs for brain-computer interface (BCI) applications. The aim of BCI is to translate brain signals into control signals for a computer (or device) to perform the desired action \cite{NicolasAlonso_2012_BCIReview}. The advancements in BCI have enabled people with neuromuscular disorders to restore or replace some of their motor functions such as limb movements \cite{Zhang_2021_BCI_motordisorderSysReview}. For a successful BCI, a user typically has to undergo training for generating brain signals that can encode their intention for communicating with the connected device. Likewise, an AI technique powering BCI also needs to be trained to decode the intention based brain signals, from the user, to command signals for successful control of the device.

The relevance of BCI for DCN studies come from gaining insights into the neural reuse of already evolved cortical circuits for performing a given function such as limb movement. Hence this re-learning of a user to control their limb via BCI instead of normal output pathways of peripheral nerves and muscles would be a key mechanism for successful BCI. In this regard, the decoding of the composition of the `control' signal, based on lower level features using multiple processing layers of CNN, can have profound implications for shedding light on the consequences of neural commitment (perceptual narrowing) for defining neural reuse. Consequently, the CNNs powering BCI can shed light on the neural reuse and perceptual narrowing to perform BCI. 
%because of the neural commitment.

In the following subsections, we review the most promising studies for both EEG-based \cite{Lawhern_2018_EEGNet} and fNIRS-based \cite{Sasikala_2020_DCN_fNIRS_BCI} BCI applications.

In general, EEG-based BCI paradigms can be categorised as 1) event-related potential (ERP) and 2) oscillation based BCI paradigms. The classic ERP based brain-computer interface (BCIs) aim to recognise a relatively high amplitude characterised with low frequency in the EEG signal evoked in response to, and time-locked with, an external event/stimulus. In contrast, the oscillation based BCI paradigms make use of the signal power pertaining to specific frequency bands for classification. A general-purpose architecture for CNN developed for the classification of EEG-based BCI paradigms, called EEGNet, is proposed by Lawhern \emph{et al.} \cite{Lawhern_2018_EEGNet}. The strength of the EEGNet lies in its successful classification for both even-related and oscillatory BCIs, as validated in the study over 4 different BCI paradigms: 3 ERP-based BCI and 1 oscillatory-based BCI.

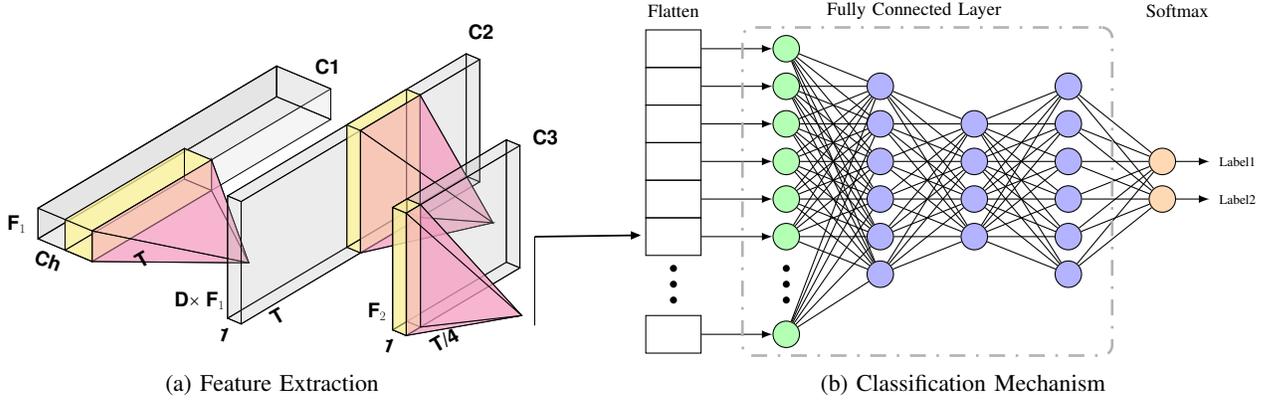
\begin{figure*}[!t]
\subfloat[Feature Extraction]{\label{Fig:CNN_FS}%{60}{40}
\tdplotsetmaincoords{60}{40}
%\begin{landscape}
%%%\begin{figure*}
%\begin{figure}
%\subfloat[][The 3D architecture of CNN]{
\scalebox{.235}{
%\vspace{-4cm}
\hspace{4cm}
\begin{tikzpicture}[tdplot_main_coords] 
\tikzstyle{outer_box} = [thick, fill=lightgray, fill opacity=.15]
\tikzstyle{inner_box} = [thick, fill=yellow!50, fill opacity=.5]
\tikzstyle{edge_fill} = [thick, fill=cerisepink!50, fill opacity=.5]
%-------------------------------------------------------
% \begin{scope}[ x={(0cm,2cm)}, y={(0cm,1.5cm))}, z={(0.5cm,1.5cm)} ]
% \node[canvas is yz plane at x=0] at (0,0) {\includegraphics[width=15cm]{Figures/EEGsignal_10.png}};
% \node[canvas is yz plane at x=1] at (0,0) {\includegraphics[width=15cm]{Figures/EEGsignal_10.png}};
% \node[canvas is yz plane at x=2] at (0,0) {\includegraphics[width=15cm]{Figures/EEGsignal_10.png}};
% \node[canvas is yz plane at x=3] at (0,0) {\includegraphics[width=15cm]{Figures/EEGsignal_10.png}};
% \end{scope}
%--------------------------------------------------------
% Outer0 bottom -- x:4, y:21, z:2 
\draw[outer_box] (-12, 0, 0) -- ++(4, 0, 0) -- ++(0, 0, 2) -- ++(-4, 0, 0) -- cycle;
% Outer1 back
\draw[outer_box] (-12, 0, 0) -- ++(4, 0, 0) -- ++(0, 21, 0) -- ++(-4, 0, 0) -- cycle;
% Outer1 left
\draw[outer_box] (-12, 0, 0) -- ++(0, 0, 2) -- ++(0, 21, 0) -- ++(0, 0, -2) -- cycle;
% Outer1 right
\draw[outer_box] (-12, 0, 0) -- ++(0, 0, 2) -- ++(0, 21, 0) -- ++(0, 0, -2) -- cycle;
% Outer1 front
\draw[outer_box] (-12, 0, 2) -- ++(4, 0, 0) -- ++(0, 21, 0) -- ++(-4, 0, 0) -- cycle;
% Outer1 top
\draw[outer_box] (-12, 21, 0) -- ++(4, 0, 0) -- ++(0, 0, 2) -- ++(-4, 0, 0) -- cycle;
% %********************************************************************

% % % Inner0 bottom -- x:2, y:10.5, z:2
\draw[inner_box] (-12+2, 0, 0) -- ++(2, 0, 0) -- ++(0, 0, 2) -- ++(-2, 0, 0) -- cycle;
% % Inner0 back
 \draw[inner_box] (-12+2, 0, 0) -- ++(2, 0, 0) -- ++(0, 10.5, 0) -- ++(-2, 0, 0) -- cycle;
% % Inner0 left
\draw[inner_box] (-12+2, 0, 0) -- ++(0, 0, 2) -- ++(0, 10.5, 0) -- ++(0, 0, -2) -- cycle;
% % Inner0 right
\draw[inner_box] (-12+2+2, 0, 0) -- ++(0, 0, 2) -- ++(0, 10.5, 0) -- ++(0, 0, -2) -- cycle;
% % Inner0 front
\draw[inner_box] (-12+2, 0, 0+2) -- ++(2, 0, 0) -- ++(0, 10.5, 0) -- ++(-2, 0, 0) -- cycle;
% % Inner0 top
 \draw[inner_box] (-12+2, 0+10.5, 0) -- ++(2, 0, 0) -- ++(0, 0, 2) -- ++(-2, 0, 0) -- cycle;
% %-------------------------------------------------
% % %-------------------------------------------------
% % % edge_0 bottom
\draw[edge_fill] (-12+4, 0, 0) -- ++(8, 4.25, 1) -- ++(-8, 6.25, -1)-- cycle ;
% % edge_0 right_behind
\draw[edge_fill] (-12+4, 0, 0) -- ++(0, 0, 2) -- ++ (8,4.25 ,-1) -- cycle;
% % edge_0 right_front
\draw[edge_fill] (-12+4, 10.5, 0) -- ++(0, 0, 2) -- ++ (8,-6.25,-1) -- cycle;
% % edge_0 top
\draw[edge_fill] (-12+4, 0, 2) -- ++(8, 4.25, -1) -- ++(-8, 6.25,1) -- cycle;
% % % %-------------------------------------------------
%********************************************************************
% Outer 0: Dimension 
%\node [\fontsize{60}{70}\selectfont Huge text, label=below:2](0, -1, 0) {};
%\node [\font=Huge, scale=10,label=below:\textbf{2}](0, -1, 0) {};
\draw (-12, 25.5, 0) node[rotate=0,font=\fontsize{32}{38}\sffamily\bfseries]{C1};
\draw (-10, -1.5, 0) node[rotate=-20,font=\fontsize{32}{38}\sffamily\bfseries]{Ch};
\draw (-14, 0.5, 0) node[rotate=0,font=\fontsize{32}{38}\sffamily\bfseries]{F\(_1\)};
\draw (-6, 2, 0) node[rotate=30,font=\fontsize{32}{38}\sffamily\bfseries]{T};
%********************************************************************
%%%%%%%%%%%%%%%%%%%%%%%%%%%%%%%%%%%%%%%%%%%%%%%%%%%%%%%%%%%%%%%%%%%%%%%%%%%%%%%%%%%%%%%%%
%--------------------------------------------------------
% Outer1 bottom -- x:1, y:21, z:8 
\draw[outer_box] (2, 0, 0) -- ++(1, 0, 0) -- ++(0, 0, 4*2) -- ++(-1, 0, 0) -- cycle;
% Outer1 back
\draw[outer_box] (2, 0, 0) -- ++(1, 0, 0) -- ++(0, 21, 0) -- ++(-1, 0, 0) -- cycle;
% Outer1 left
\draw[outer_box] (2, 0, 0) -- ++(0, 0, 4*2) -- ++(0, 21, 0) -- ++(0, 0, -4*2) -- cycle;
% Outer1 right
\draw[outer_box] (2+1, 0, 0) -- ++(0, 0, 4*2) -- ++(0, 21, 0) -- ++(0, 0, -4*2) -- cycle;
% Outer1 front
\draw[outer_box] (2, 0, 4*2) -- ++(1, 0, 0) -- ++(0, 21, 0) -- ++(-1, 0, 0) -- cycle;
% Outer1 top
\draw[outer_box] (2, 21, 0) -- ++(1, 0, 0) -- ++(0, 0, 4*2) -- ++(-1, 0, 0) -- cycle;
%********************************************************************
% Outer 1: Dimension 
%\node [\fontsize{60}{70}\selectfont Huge text, label=below:2](0, -1, 0) {};
%\node [\font=Huge, scale=10,label=below:\textbf{2}](0, -1, 0) {};
\draw (1, 23.5, 8) node[rotate=0,font=\fontsize{32}{38}\sffamily\bfseries]{C2};
\draw (3, -1.5, 0) node[rotate=-20,font=\fontsize{32}{38}\sffamily\bfseries]{1};
\draw (-0.5, 0.5, 0) node[rotate=0,font=\fontsize{32}{38}\sffamily\bfseries]{D\(\times\) F\(_1\)};
\draw (4, 2, 0) node[rotate=30,font=\fontsize{32}{38}\sffamily\bfseries]{T};
%********************************************************************

% % Inner1 bottom -- x:2, y:10.5, z:8
\draw[inner_box] (2, 10.5, 0) -- ++(1, 0, 0) -- ++(0, 0, 4*2) -- ++(-1, 0, 0) -- cycle;
% % Inner1 back
 \draw[inner_box] (2, 10.5, 0) -- ++(1, 0, 0) -- ++(0, 10.5/2, 0) -- ++(-1, 0, 0) -- cycle;
% % Inner1 left
\draw[inner_box] (2, 10.5, 0) -- ++(0, 0, 4*2) -- ++(0, 10.5/2, 0) -- ++(0, 0, -4*2) -- cycle;
% % Inner1 right
\draw[inner_box] (2+1, 10.5, 0) -- ++(0, 0, 4*2) -- ++(0, 10.5/2, 0) -- ++(0, 0, -4*2) -- cycle;
% % Inner1 front
\draw[inner_box] (2, 10.5, 0+4*2) -- ++(1, 0, 0) -- ++(0, 10.5/2, 0) -- ++(-1, 0, 0) -- cycle;
% % Inner1 top
 \draw[inner_box] (2, 10.5+10.5/2, 0) -- ++(1, 0, 0) -- ++(0, 0, 4*2) -- ++(-1, 0, 0) -- cycle;
%-------------------------------------------------
% %-------------------------------------------------
% % edge_1 bottom
\draw[edge_fill] (1+2, 10.5, 0) -- ++(8, 4.25/2, 2*2) -- ++(-8, 6.25/2, -2*2)-- cycle ;
% % edge_1 right_behind
\draw[edge_fill] (1+2, 10.5, 0) -- ++(0, 0, 4*2) -- ++ (8,4.25/2 ,-2*2) -- cycle;
% % edge_1 right_front
\draw[edge_fill] (1+2, 10.5+10.5/2, 0) -- ++(0, 0, 4*2) -- ++ (8,-6.25/2,-2*2) -- cycle;
% % edge_1 top
\draw[edge_fill] (1+2, 10.5, 4*2) -- ++(8, 4.25/2, -2*2) -- ++(-8, 6.25/2,2*2) -- cycle;
% % %-------------------------------------------------
% \draw[edge_fill] (4, 10.5, 0) -- ++(6, 0, 0) -- ++ (-6,10.5,0)-- cycle ;
% % % edge_1 right_behind
% \draw[edge_fill] (4, 10.5, 0) -- ++(0, 0, 2) -- ++ (6,0 ,-2) -- cycle;
% % % edge_1 right_front
% \draw[edge_fill] (4, 21, 0) -- ++(0, 0, 2) -- ++ (6,-10.5,-2) -- cycle;
% % % edge_1 top
% \draw[edge_fill] (4, 10.5, 2) -- ++(0, 10.5, 0) -- ++(6,-10.5,-2) -- cycle;
% % 
%-------------------------------------------------
% %-------------------------------------------------
%% distance from outer 1 box: 8
% % Outer2 bottom -- x:2, y:10, z:4 
\draw[outer_box] (2+8, 5, 0) -- ++(1, 0, 0) -- ++(0, 0, 4*2) -- ++(-1, 0, 0) -- cycle;
% % Outer2 back
\draw[outer_box] (2+8, 5, 0 ) -- ++(1, 0, 0) -- ++(0, 10, 0) -- ++(-1, 0, 0) -- cycle;
% % Outer2 left
\draw[outer_box] (2+8, 5, 0) -- ++(0, 0, 4*2) -- ++(0, 10, 0) -- ++(0, 0, -4*2) -- cycle;
% % Outer1 right
 \draw[outer_box] (2+8+1, 5, 0) -- ++(0, 0, 4*2) -- ++(0, 10, 0) -- ++(0, 0, -4*2) -- cycle;
% % Outer1 front
\draw[outer_box] (2+8, 5, 0+4*2) -- ++(1, 0, 0) -- ++(0, 10, 0) -- ++(-1, 0, 0) -- cycle;
% % Outer1 right
 \draw[outer_box] (2+8, 5+10, 0) -- ++(1, 0, 0) -- ++(0, 0, 4*2) -- ++(-1, 0, 0) -- cycle;
 
%********************************************************************
% Outer 2: Dimension 
%\node [\fontsize{60}{70}\selectfont Huge text, label=below:2](0, -1, 0) {};
%\node [\font=Huge, scale=10,label=below:\textbf{2}](0, -1, 0) {};
\draw (7, 22, 4) node[rotate=0,font=\fontsize{32}{38}\sffamily\bfseries]{C3};
\draw (2+8+1, 5-1.5, 0) node[rotate=-20,font=\fontsize{32}{38}\sffamily\bfseries]{1};
\draw (4+8+1, 6, 0) node[rotate=30,font=\fontsize{32}{38}\sffamily\bfseries]{T/4};
\draw (8, 6, 0) node[rotate=0,font=\fontsize{32}{38}\sffamily\bfseries]{F\(_2\)};
%********************************************************************

% % Inner2 bottom -- x:1, y:5, z:4
\draw[inner_box] (2+8, 5, 0) -- ++(1, 0, 0) -- ++(0, 5/4, 0) -- ++(-1, 0, 0) -- cycle;
% % % Inner2 back
\draw[inner_box] (2+8, 5, 0) -- ++(0, 0, 4*2) -- ++(0, 5/4, 0) -- ++(0, 0, -4*2) -- cycle;
% % % Inner2 left
\draw[inner_box] (2+8, 5, 0) -- ++(1, 0, 0) -- ++(0, 0, 4*2) -- ++(-1, 0, 0) -- cycle;
% % % Inner2 right
\draw[inner_box] (2+8, 5+5/4, 0) -- ++(1, 0, 0) -- ++(0, 0, 4*2) -- ++(-1, 0, 0) -- cycle;
% % % Inner2 front
\draw[inner_box] (2+8+1, 5, 0) -- ++(0, 0, 4*2) -- ++(0, 5/4, 0) -- ++(0, 0, -4*2) -- cycle;
% % % Inner2 top
\draw[inner_box] (2+8, 5, 0+4*2) -- ++(1, 0, 0) -- ++(0, 5/4, 0) -- ++(-1, 0, 0) -- cycle;

% %-------------------------------------------------
% % edge_2 bottom
\draw[edge_fill] (2+8+1, 5, 0) -- ++(8, 2.5/4, 2*2) -- ++ (-8, 2.5/4,-2*2) -- cycle;
% % % edge_2 right_front
\draw[edge_fill] (2+8+1, 5, 4*2) -- ++(8, 2.5/4, -2*2) -- ++ (-8, -2.5/4,-2*2) -- cycle;
% % % % edge_2 right_behind
\draw[edge_fill] (2+8+1, 5+5/4, 4*2) -- ++(8, -2.5/4, -2*2) -- ++ (-8, 2.5/4,-2*2) -- cycle;
% % % edge_1 top
\draw[edge_fill] (2+8+1, 5, 4*2) -- ++(8, 2.5/4, -2*2) -- ++(-8, 2.5/4, 2*2) -- cycle;
\draw[ultra thick, - ] (16.5, 9.75, 6.4) -- ++(0, 0, -5.8);
\draw[ultra thick, ->, -triangle 45,
        line width=1mm] (16.5, 9.75, 6.4) -- ++(7, 1, 2.25);
\end{tikzpicture}}} \hspace{-.4cm}
\subfloat[Classification Mechanism]{\label{Fig:CNN_CM}\begin{tikzpicture}[>=latex]
\hspace{0cm}
%-------------------------------------------
\tikzstyle{block} = [rectangle, draw, text width=.5cm, text badly centered, minimum height = .5 cm, font=\tiny ]
\tikzstyle{neuron} = [circle, draw, inner sep = 0cm, font= \tiny,  minimum size=10pt]
\tikzstyle{dotted_block} = [draw=black!30!white, line width=1pt, dash pattern=on 1pt off 4pt on 6pt off 4pt,
            inner ysep=1mm,inner xsep=4mm, rectangle, rounded corners ]
%----------------------------------------------
\foreach \x in {0}
    \foreach \y in {0,...,5} 
      {\pgfmathtruncatemacro{\label}{\x - 5 *  \y +10}
      \node (f\x\y) [block] at (\x,.5*\y) {};}
% Flatten layer       
\node[block, draw = none, rotate = 90, font = \Huge] at (0, -.7) {...};
\node(f06)[block] at (0, -1.3) {};
%-------------------------------------------
%-------------------------------------------
% Neural Network 
%-------------------------------------------
%-------------------------------------------
% input layer
%-------------------------------------------
\foreach \n in {0,...,5}
    {\node (n\n) [neuron,fill = green!30, right of = f0\n, node distance = 1.5cm] {};} 
\node (n_dots)[block, draw = none, rotate = 90, font = \Huge] at (1.5, -.7) {...};
\node (n6) [neuron, right of = f06,fill = green!30, node distance = 1.5cm] {};
%-------------------------------------------
% hidden layer1
%-------------------------------------------
\foreach \n in {0,...,4}
    {\node (h1\n) [neuron, right of = n\n, fill = blue!30, node distance = 1.25cm] {};} 
\node (h15) [neuron, below of = h10, node distance = .5cm, fill = blue!30] {};
%-------------------------------------------
% hidden layer2
%-------------------------------------------
\foreach \n in {0,...,3}
    {\node (h2\n) [neuron, right of = h1\n , fill = blue!30,  node distance = 1.25cm] {};} 
%-------------------------------------------
% hidden layer3
%-------------------------------------------
\foreach \n in {0,...,4}
    {\node (h3\n) [neuron, right of = h1\n,fill = blue!30,  node distance = 2.5cm] {};} 
\node (h35) [neuron, below of = h30, fill = blue!30,node distance = .5cm] {};
% output layer
\foreach \n in {0,...,1}
    {\node (o\n) [neuron, right of = h3\n, yshift = .5cm, node distance = 1.25cm, fill = orange!30] {};} 
% label the output nodes with class
\node (c0)[block, draw =none, right of = o0] {Label2};
\node (c1)[block, draw =none, right of = o1] {Label1};
%---------------------------------------
% paths of neural networks
%---------------------------------------
% fully connected to input layer
\foreach \i  in {0,...,6} 
    {\draw [->]  (f0\i)-- (n\i);}
% input to hidden layer 1
\foreach \n  in {0,...,6} 
    \foreach \h in {0,...,5}
    {\draw [-]  (n\n)-- (h1\h);}
% hidden layer 1 to hidden layer 2
\foreach \h  in {0,...,5} 
    \foreach \i in {0,...,3}
    {\draw [-]  (h1\h)-- (h2\i);}
% hidden layer 2 to hidden layer 3
\foreach \h  in {0,...,3} 
    \foreach \i in {0,...,5}
    {\draw [-]  (h2\h)-- (h3\i);}
% hidden layer 3 to output
\foreach \h  in {0,...,5} 
    \foreach \i in {0,...,1}
    {\draw [-]  (h3\h)-- (o\i);}
% output layer to labels
\foreach \i  in {0,...,1} 
    {\draw [->]  (o\i)-- (c\i);}
%----------------------------------------------------
%name the layers
\node[block, draw=none, text width = 1cm, font= \scriptsize] at (0, 3) {Flatten};
\node[block, draw=none, text width = 3cm, font= \scriptsize] at (3.2, 3) {Fully Connected Layer};
\node[block, draw=none, text width = 3cm, font= \scriptsize] at (6.7, 3) {Softmax};
\node[dotted_block, fit = (n0)(n5)(n6)(h10)(h15)(h20)(h23)(h30)(h35)] {};
\end{tikzpicture}} \hspace{.1cm}
\caption{A schematic of the EEGNet \cite{Lawhern_2018_EEGNet} architecture that gave the best classification results over 4 different brain computer interface (BCI) EEG based datasets. (a) The first stage consists of a series of convolutional (C) and pooling (P) layers of varying dimensions where F\(_1\) and F\(_2\) denote respective filter sizes, Ch is the number of electrode channels and T stands for the number of time points. D represents the number of spatial filters. F\(_1\) is a temporal filter, whereas F\(_2\) is a pointwise filter. A feature map is constructed from the input matrix by convoluting it with a filter (or kernel). By using different filters on the same input, different features from the input image can be detected such as an edge. To account for nonlinearities, the feature map is then passed through an activation function. The pooling layer is used to merge semantically similar features found from convolution layers into one. The most common pooling function used is the \emph{max} pooling but in this study \cite{Lawhern_2018_EEGNet} average pooling is used. (b) The feature set thus computed, through convolution and pooling layers, is then flattened and input to a fully connected layer (a neural network). The softmax layer, which is also an activation function, is the last layer that finally predicts the label for the class. }
\label{fig:EEGNet_architecture}
\end{figure*}

The proposed architecture of EEGNet is illustrated in Fig. \ref{fig:EEGNet_architecture}. In reference to Fig. \ref{fig:EEGNet_architecture}, EEGNet undertakes the following convolutions to learn respective lower level features from the input EEG data:
\begin{enumerate}[label=(\roman*)]
    \item \emph{C1}: Temporal convolution to learn frequency filters. 
    \item \emph{C2}: Depth-wise convolution to learn frequency-specific spatial filters (i.e.\textcolor{blue}{,} a specific spatial filter for each frequency filter).
    \item \emph{C3}: Separable convolution to optimally aggregate the features maps together.
\end{enumerate}

%------------------------------------------------------
% -------------  CNN Architecture 
%----------------------------------------------------
%------------------------------------------------------
% -------------------- END
% -------------  CNN Architecture 
%----------------------------------------------------

Since EEGNet is tested on 4 different EEG datasets, Ch is used to denote the number of electrodes and T represents the time samples for a particular dataset. The number of hyperparameters, per BCI paradigm for a total of 4 paradigms investigated (P1-P4), to be learnt by EEGNet with 4 temporal filters and 2 spatial filters per temporal filter, are: [P1: 1066, P2: 1082, P3: 1098, and P4: 796]. In general, the performance of EEGNet was superior for ERP-based BCI paradigms in comparison to oscillatory-based BCI with an average of \(\sim\) 80\% classification accuracy across the 4 paradigms.

%feature interpretation
In an attempt to validate that their proposed EEGNet model learning is based on relevant features depicting brain activity, the authors investigated three different approaches for enabling feature explainability:
\begin{enumerate}[label=(\roman*)]
    \item \emph{Hidden unit activations}: This was done after depth wise convolution: i.e.\textcolor{blue}{,} C2 in Fig. \ref{fig:EEGNet_architecture} shed\textcolor{blue}{s} light on the spatial localisation of the activations corresponding to a particular frequency.
    \item \emph{Filter weights:} The visualisation of filter weights was possible because of EEGNets architecture that limits the connectivity between two convolution layers: i.e., direct visualisation of the narrow band filter frequency filters weight for C1, and the frequency-specific spatial filter weights for C2 in Fig. \ref{fig:EEGNet_architecture} sheds light on the relevant frequency components, and frequency specific spatial localisation.
    \item \emph{Feature relevance:} The relevance of individual features for classification performance of EEGNet was calculated on a per trial basis using DeepLIFT algorithm. 
\end{enumerate}

The validity of the features, on whose basis is the inference mechanism learnt, is of significance to establish the robustness of the CNN architecture. However, the almost \(\sim\) 950 learnt hyperparameters are not explainable since a given optimised value of a hyperparameter can not be matched to a particular representation of brain activity. In essence, the optimal values of the hyperparameters of EEGNet are a filter that can not shed light on the interconnections or EC of cortical regions.

%--------------------------------------------------------
\noindent\emph{e) RepL with fNIRS using CNN}
%--------------------------------------------------------

The classic analysis paradigms for fNIRS signals are based on the statistical features most representative of the underlying activity. For representation learning on fNIRS signals using CNNs, the fNIRS signals are first transformed to equivalent image time-frequency representations known as spectrograms.

In the work by Janani \emph{et al.} \cite{Sasikala_2020_DCN_fNIRS_BCI} the authors investigated the possibility of classification of four different motor imagery tasks, i.e., participants \emph{imagined} moving their limbs instead of physically moving their limbs, using CNNs. More specifically, the four different motor imagery tasks were: right- and left-fist clenching, right- and left-foot tapping. The fNIRS channels were placed on top of the left and right hemispheres to record brain activity from respective cortical regions.

The spectrogram method was used to transform the fNIRS signal into a time-frequency image. The architecture of the CNN feature extraction stage had two convolution (C) layers with 1 pooling (P) layer in between the following sizes- C1: 3\(\times\)23\(\times\)16; P: 2\(\times\)12\(\times\)16 and C2: 3\(\times\)3\(\times\)32. The fully connected layer, had 288 nodes which connected through hidden layers, classified the input fNIRS image into 4 motor imagery tasks.  

The average classification accuracy obtained over all four tasks was 72.35\%. Although the CNN preformed the best amongst other standard AI methods (SVM and multi-layer perceptron), the classification accuracy was not at par with the usual high performing CNNs. The modest performance of CNN could be attributed to the large input fNIRS image dimensions (660\(\times\)22). 

%------------------------------------------------------
\noindent\emph{f) MVPA with EEG using SVM} \label{sec:MVPA_EEG_SVM}
%------------------------------------------------------

The MVM for EEG signals can be built using ERPs, wavelet coefficients or using component analysis. For ERPs, taking the average is beneficial for reducing some noise though single ERPs are more representative. In addition, if the MVPA investigation with respect to the cortical regions of the brain is critical, then source localisation of the electrodes is important.

A toolbox that has been designed in particular to make the MVPA more accessible is the Amsterdam Decoding and Modeling Toolbox (ADAM) \cite{Fahrenfort_2018_ADAMtoolbox}. It takes as input EEG data in standard formats of FileldTrip or EEGLAB and is able to pre-process i.e., increase signal to noise ratio, remove motion artefacts. The first level MVPA can compute a performance metric, whereas group level MVPA can compute statistical significance for patterns.

A successful application of MVPA with EEG data for the detection of a face in the wild (natural settings) is done by Cauchoix \emph{et al.} \cite{Cauchoix_2014_NeuralDynamicsOfFaceDetection}. The stimulus images were grayscale photographs of human faces presented in their natural contexts.

After pre-processing the EEG signals, ERPs were computed separately for correct human face target trials and correct animal face non-target trials. For face processing in adult EEG data, MVPA has been successfully used for face detection in wild settings for ERPs: P100 (positive potential observed after around 100msec of stimulus presentation) at four bilateral occipital electrodes (O1, O2, PO3, and PO4) and for the N170, (negative potential observed after around 170msec of stimulus presentation) at four right hemisphere occipitotemporal electrodes (PO10, PO8, P8, and TP8). 

Their MVPA results with SVM achieved a classification accuracy of 94.8\%. In addition, based on their results, they suggest that neural dynamics of face detection could be readout very early, starting \(\sim\)95 ms following stimulus onset. 

Although the above reviewed works involving SVM, i.e. Moezzi \emph{et al.}  \cite{Moezzi_2019_EEG_SVM_FC} (section \ref{sec:NonXAI_Adults}(a)) and Cauchoix \emph{et al.} \cite{Cauchoix_2014_NeuralDynamicsOfFaceDetection} (section \ref{sec:NonXAI_Adults}(f)) are non-explainable methods, other works involving SVMs in particular have investigated gaining an insight into the inference mechanism by using logic programming \cite{Shakerin_2020}, and decision trees \cite{Vieira_2020}. Given the aforementioned limitations of non-explainable AI methods reviewed to inform the underlying brain mechanisms, we review the XAI methods in the next section.

%----------------------------------------------------------
\subsubsection{XAI Methods}
%----------------------------------------------------------

In this section, we review the XAI methods, i.e., AI methods whose inference mechanism can be explained in terms of the brain activity patterns. In addition, the insights obtained owing to the explainability of the applied XAI method for the given task are also investigated.

%The XAI methods reviewed are as follows:
% \begin{enumerate}[label=(\alph*)]
%     \item EC with fNIRS using Effective Fuzzy Cognitive Maps (EFCMs)
%     \item RepL with EEG using independent component analysis and Fuzzy Neural Networks (ICA-FNNs)
% \end{enumerate}

%----------------------------------------------------------
\noindent\emph{a) EC with fNIRS using Effective Fuzzy Cognitive Maps (EFCMs)} 
%----------------------------------------------------------

An fNIRS study that estimated EC amongst fNIRS channels (corresponding anatomical regions in the cortex) based on fuzzy cognitive maps (FCMs) is proposed by Kiani \emph{et al.} \cite{Kiani_2019_EFCM}. A FCM is a cognitive mapping technique based on graph theory, with a formal mathematical definition given as follows in (\ref{Eq:FCM}).

\begin{equation} \label{Eq:FCM}
    C_{j(t+1)} = f(\sum_{i=1}^{N}e_{ij}C_i(t))
\end{equation}

where \(N\) is the number of concepts (or fNIRS channels) in a given system, \(C_j(t)\) is the value of a given concept \(C_j\) at iteration \(t\), \(e_{ij}\) are the fuzzy weights or EC that concept \(C_i\) exerts on concept \(C_j\) and \(f\) is typically a sigmoid function that scales the weights to [-1,1] for comparative analysis such that a value of 1 means fully interconnected, a value of -1 means  fully  interconnected  in  the  opposite  direction, a value of 0 means disconnected, and a value between 0 and 1 (or -1) means interconnected to a certain extent. The optimal values of EC (\(e_{ij}\)) are typically found using an evolutionary algorithm such as Genetic algorithm (GA) (GA has also been used in the aforementioned study \cite{Kiani_2019_EFCM}).

%-----------------------------------------------------------
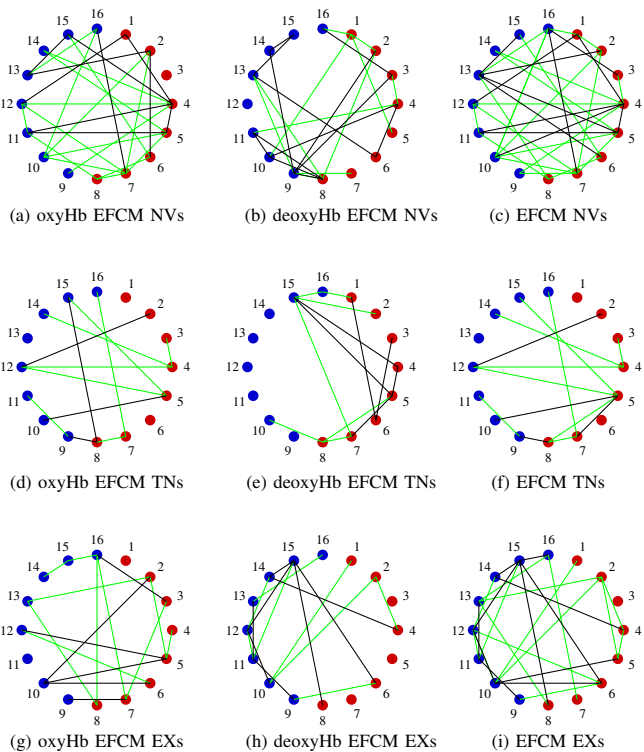
\begin{figure}[!h] % L B R T -- trim arguments
\centering
%----------------------------------------------------------
%\includegraphics[scale=0.5, trim={1.5cm 7.35cm 1.5cm 1.5cm}, clip = true]{Figures/Fig9.pdf}
%----------------------------------------------------------
\scalebox{1}{\tikzstyle{ch_PFC} = [circle, draw, fill=orange, 
    text width=1cm, text centered, rounded corners, minimum size=.8cm]
\tikzstyle{ch_MC} = [circle, draw, fill=brown, 
    text width=1cm, text centered, rounded corners, minimum size=.8cm, opacity=1]
\tikzstyle{dashdotted}= [dash pattern={on 5pt off 3pt on 2pt off 3pt}]
\tikzset{>={Latex[width=1mm,length=2mm]}}
%---------------------------------------------------------------
%trim={<left> <lower> <right> <upper>}
\begin{tikzpicture}[node distance =1.5cm, baseline]
\hspace{-0.5cm}
%-----------------------------------------------------------
%----------------- NVs --------------------
%-----------------------------------------------------------
% \node (N_ch1)[ch_PFC, text badly centered, minimum size=.9cm, scale = 0.5]{\Huge \textbf{1}};
% \node (N_ch2)[ch_PFC, text badly centered, minimum size=.9cm, scale = 0.5, below of = N_ch1, xshift = .5cm]{\Huge \textbf{2}};
% %-----------------------------------------------------------
% \node (N_ch8)[ch_MC, text badly centered, minimum size=.9cm, scale = 0.5, below of=N_ch1, node distance = 4cm]{\Huge \textbf{8}};
%-----------------------------------------------------------
%paths
% \draw [->, cyan, arrows={-Triangle[angle=90:4pt,black,fill=black]},line width=.5mm, solid] (13.7, .8) -- (13.5, .3); % R2 - R3
%-----------------------------------------------------------
%\draw(0,0) circle (1);
\pgfmathsetmacro\n{16}
\pgfmathsetmacro\c{3}
% \foreach \i/\k in {0/A,1/B,2/C,3/X,4/Y,5/Z} {
%     \pgfmathsetmacro\r{\i*(360/\n)}
%     \fill[red] (\r:1) circle (2pt) coordinate (n-\i);
%     \node at (\r:1.2) {\k};
% }
% \draw (n-0) -- (n-3);
% \draw (n-0) -- (n-5);
% \draw (n-0) -- (n-2);
%---------------------------------------
% Novices
%---------------------------------------
\foreach \c in {0, 3, 6}{ 
%\draw(\c,0) circle (1);
\foreach \i in {0, 1, 2, ..., 15} {
    \pgfmathsetmacro\r{\i*(360/\n)+90}
    \pgfmathsetmacro\l{int(\n -\i)}
    \ifnum\i<8
        \fill[blue!80!black, xshift = \c cm] (\r:1) circle (2pt) coordinate (n\c\i);
    \else
        \fill[red!80!black,  xshift = \c cm] (\r:1) circle (2pt) coordinate (n\c\i);
    \fi
    \node at ([xshift = \c cm]\r:1.2) {\tiny \l};
}
}
% captions
\node(label_NVs1) at (0,-1.5)[draw = none]{\scriptsize (a) oxyHb EFCM NVs};
\node(label_NVs2) at (3.25,-1.5)[draw = none]{\scriptsize (b) deoxyHb EFCM NVs};
\node(label_NVs3) at (6,-1.5)[draw = none]{\scriptsize (c) EFCM NVs};
%---------------------------------------
% Trainees
%---------------------------------------
\foreach \c in {0, 3, 6}{ 
%\draw(\c,-3.5) circle (1);
\foreach \i in {0, 1, 2, ..., 15} {
    \pgfmathsetmacro\r{\i*(360/\n)+90}
    \pgfmathsetmacro\l{int(\n -\i)}
    \ifnum\i<8
        \fill[blue!80!black, xshift = \c cm, yshift =-3.5cm] (\r:1) circle (2pt) coordinate (t\c\i);
    \else
        \fill[red!80!black,  xshift = \c cm, yshift =-3.5cm] (\r:1) circle (2pt) coordinate (t\c\i);
    \fi
    \node at ([xshift = \c cm, , yshift =-3.5cm]\r:1.2) {\tiny \l};
}
}
\node(label_TNs1) at (0,-5.05)[draw = none]{\scriptsize (d) oxyHb EFCM TNs};
\node(label_TNs2) at (3.25,-5.05)[draw = none]{\scriptsize (e) deoxyHb EFCM TNs};
\node(label_TNs3) at (6,-5.05)[draw = none]{\scriptsize (f) EFCM TNs};
%-----------------------------------------------------
%---------------------------------------
% Experts
%---------------------------------------
\foreach \c in {0, 3, 6}{ 
%\draw(\c,-7) circle (1);
\foreach \i in {0, 1, 2, ..., 15} {
    \pgfmathsetmacro\r{\i*(360/\n)+90}
    \pgfmathsetmacro\l{int(\n -\i)}
    \ifnum\i<8
        \fill[blue!80!black, xshift = \c cm, yshift =-7cm] (\r:1) circle (2pt) coordinate (e\c\i);
    \else
        \fill[red!80!black, xshift = \c cm, yshift =-7cm] (\r:1) circle (2pt) coordinate (e\c\i);
    \fi
    \node at ([xshift = \c cm, , yshift =-7cm]\r:1.2) {\tiny \l};
}
}
\node(label_EXs1) at (0,-8.5)[draw = none]{\scriptsize (g) oxyHb EFCM EXs};
\node(label_EXs2) at (3.25,-8.5)[draw = none]{\scriptsize (h) deoxyHb EFCM EXs};
\node(label_EXs3) at (6,-8.5)[draw = none]{\scriptsize (i) EFCM EXs};
%-------------------------------------------
% Paths
%------------------------------------------------------
%Novices Oxy
%pos
\foreach \i/\j in{1/10,2/11,2/12,3/0,4/5,4/12,4/9,5/8, 6/0,6/14,6/11,6/9,7/12,8/9,8/10,9/11,9/14}{
    \draw [- , green] (n0\i) -- (n0\j);
}
%neg
\foreach \i/\j in{1/3,1/12,3/14,4/15,5/11,5/12,9/0,10/14,11/12,12/15}{
    \draw [- , black] (n0\i) -- (n0\j);
}
%Novices DeOxy
%pos
\foreach \i/\j in{3/7,3/8,5/12,6/15,7/8,7/9,11/15,8/14,12/13,14/15,0/15}{
    \draw [- , green] (n3\i) -- (n3\j);
}
%neg
\foreach \i/\j in{1/2,1/3,2/7,3/10,5/6,5/8,6/8,6/12,7/14,7/13,7/8,10/12,13/0}{
    \draw [- , black] (n3\i) -- (n3\j);
}
%Novices Eff
%pos
\foreach \i/\j in{0/3,0/6,0/15,1/10,2/11,2/12,3/8,4/12,4/9,4/5,5/8,6/0,6/14,6/11,6/9,7/12,7/9,8/10,9/11,9/14,12/13,14/15}{
    \draw [- , green] (n6\i) -- (n6\j);
}
%neg
\foreach \i/\j in{0/9,0/13,1/3,3/11,3/14,4/15,3/10,5/11,5/12,6/12,11/12,12/15}{
    \draw [- , black] (n6\i) -- (n6\j);
}
%------------------------------------------------------
%TNs  Oxy
%pos
\foreach \i/\j in{0/9,1/11,2/12,4/11,4/12,5/7,8/9,12/13}{
    \draw [- , green] (t0\i) -- (t0\j);
}
%neg
\foreach \i/\j in{1/8,4/14,6/11,7/8}{
    \draw [- , black] (t0\i) -- (t0\j);
}
%------------------------------------------------------
%TNs  deOxy
%pos
\foreach \i/\j in{0/1,0/15,1/14,1/9,6/8,8/11,8/9}{
    \draw [- , green] (t3\i) -- (t3\j);
}
%neg
\foreach \i/\j in{1/11,1/12,9/11,10/15,10/13,11/12}{
    \draw [- , black] (t3\i) -- (t3\j);
}
%------------------------------------------------------
%TNs  effective
%pos
\foreach \i/\j in{0/9, 1/11, 2/12, 4/11, 4/12, 5/7, 8/9, 8/11, 12/13}{
    \draw [- , green] (t6\i) -- (t6\j);
}
%neg
\foreach \i/\j in{4/14,6/11,7/8,9/11}{
    \draw [- , black] (t6\i) -- (t6\j);
}
%------------------------------------------------------
%EXs  Oxy
%pos
\foreach \i/\j in{0/1, 0/8,0/9,1/2,3/14,3/8,4/10,9/13,11/12,11/14}{
    \draw [- , green] (e0\i) -- (e0\j);
}
%neg
\foreach \i/\j in{0/13,4/11,6/10,6/11,6/14,7/9}{
    \draw [- , black] (e0\i) -- (e0\j);
}
%EXs  deOxy
%pos
\foreach \i/\j in{0/3,1/5,2/4,4/5,6/14,6/15,7/10,12/14}{
    \draw [- , green] (e3\i) -- (e3\j);
}
%neg
\foreach \i/\j in{1/2,1/4,1/8,1/10,2/12,3/5,5/7,4/6}{
    \draw [- , black] (e3\i) -- (e3\j);
}
%EXs  eff
%pos
\foreach \i/\j in{0/9, 0/3, 1/5, 2/4, 3/14, 3/8, 4/10, 4/5,4/6,6/15,6/14,7/10,11/14,11/12,12/14}{
    \draw [- , green] (e6\i) -- (e6\j);
}
%neg
\foreach \i/\j in{0/1,1/2,1/4,1/8,1/10,2/12,3/5,4/6,5/7,6/10,6/11}{
    \draw [- , black] (e6\i) -- (e6\j);
}

% \draw(3,0) circle (1);
% \pgfmathsetmacro\n{16}
% \foreach \i in {0, 1, 2, ..., 15} {
%     \pgfmathsetmacro\r{\i*(360/\n)+90}
%     \pgfmathsetmacro\l{int(\n -\i)}
%     \ifnum\i<8
%         \fill[yellow, xshift = 3cm] (\r:1) circle (2pt) coordinate (n-\i);
%     \else
%         \fill[pink, xshift = 3cm] (\r:1) circle (2pt) coordinate (n-\i);
%     \fi
%     \node at ([xshift = 3cm]\r:1.2) {\tiny \l};
% }
% %---------------------------------------
% %---------------------------------------
% \draw(6,0) circle (1);
% \pgfmathsetmacro\n{16}
% \foreach \i in {0, 1, 2, ..., 15} {
%     \pgfmathsetmacro\r{\i*(360/\n)+90}
%     \pgfmathsetmacro\l{int(\n -\i)}
%     \ifnum\i<8
%         \fill[yellow, xshift = 6cm] (\r:1) circle (2pt) coordinate (n-\i);
%     \else
%         \fill[pink, xshift = 6cm] (\r:1) circle (2pt) coordinate (n-\i);
%     \fi
%     \node at ([xshift =6cm]\r:1.2) {\tiny \l};
% }
%---------------------------------------
%---------------------------------------
\end{tikzpicture}}
%--------------------------------
\caption{(a) The Effective Connectivity (EC) networks, as delineated by the work of Kiani \emph{et al.} \cite{Kiani_2019_EFCM} for oxyHb and deoxyHb fNIRS signals recorded from prefrontal cortex (PFC), denoted in red, and motor cortex (MC), denoted in blue, of surgeons with varying levels of expertise in performing a complex visual-spatial task (more specifically laparoscopic  surgery(LS)). The expertise level of the participating subjects was categorised into three levels of Novices (NVs), Trainees (TNs), and Experts (EXs). The aim of the study was to discern the difference in EC networks formed with varying levels of proficiency for carrying out a task that requires active planning and visual-motor coordination. A green line signifies the presence of a positive (reinforcement) EC between the connecting cortical regions and the presence of a black line denotes a negative (weaken) EC between the connecting cortical regions.}
\label{fig:EFCMresults}
\end{figure}
%-----------------------------------------------------------
%-----------------------------------------------------------

The error between the estimated signal and real signal using the learnt EC weights by EFCM is computed using eq (\ref{eq:EFCMerror})

\begin{equation} \label{eq:EFCMerror}
    error=\sum_t^{T}\sum_i^{N}|C_i(t) - \hat{C}_i(t)|
\end{equation}

The proposed FCM in Kiani \emph{et al.} \cite{Kiani_2019_EFCM} is an enhanced FCM, called effective FCM (EFCM), that optimises the strength (scalar magnitude without direction) and direction separately, rendering the EFCM with more degrees of freedom to find the optimum values of EC i.e., \(e_{ij}\) in (\ref{Eq:FCM}). In addition, they also propose tuning of the transformation (sigmoid) function, \(f\) in (\ref{Eq:FCM}), to optimise how fast the non-normalised fuzzy degrees of relationship are squeezed into the normalised range for the fuzzy degrees of relationship. 

They applied their proposed EFCM on a neuroergonomics study in which brain activity of subjects, with three varying levels of expertise in performing a surgical task, was recorded using fNIRS channel placed on the prefrontal cortex (PFC) and motor cortex (MC). The EC networks found using EFCM are shown in Fig. \ref{fig:EFCMresults}. They reported an error of 120.7, as defined in (\ref{eq:EFCMerror}), averaged for all three levels of expertise in regressing the EC. 

%\subsection{XAI model of EC using EFCMs}
%--------------------------------------------
% EFCM
The EFCMs propose a partial explainable model in terms of estimating the EC as fuzzy weights (i.e., \( e_{ij}\)) between its concepts (fNIRS channels) which can be readily mapped to anatomical locations in the brain. In the original study of EFCMs \cite{Kiani_2019_EFCM} EC was estimated separately for subjects with varying levels of expertise, as shown in Fig. \ref{fig:EFCMresults}. Hence the derived EC could shed light on how the cortical networks differ in their influence on each other to subserve the complex visual-motor task on skills acquisition. In this regard EFCMs, when applied to DCN studies for estimating EC, can shed light on how developing cortical networks change in terms of their influence (EC) on account of specialisation and neural reuse processes performing a certain task.

In addition to estimating EC with statistical significance, EFCM work \cite{Kiani_2019_EFCM} also demonstrated the prowess to analyse the difference between estimating EC from oxyHb and deoxyHb dimensions of fNIRS signals for representing the EC in the cortical networks. Although it remains to be established which dimension of fNIRS is more representative for a certain task or specialisation level, they proposed that EC estimated using deoxyHb is more representative of the underlying EC as an individual gains experience in a certain motor task. %Although it warrants further investigation for determining if deoxyHb fNIRS signal becomes more representative in general on account of specialisation, in the context of the specialisation of developing brain future longitudinal DCN studies may also investigate this.

%----------------------------------------------------------
\noindent\emph{b) RepL with EEG using independent component analysis and Fuzzy Neural Networks (ICA-FNNs)}
%------------------------------------------------

In the work by Lin \emph {et al.} \cite{CTLin_2006_ICAFNN}, the cognitive state of individuals while driving in a virtual-reality based driving environment, is measured using an EEG-based XAI method. In particular, their adaptive method for recognition of drowsiness of an individual is based on a combination of independent component analysis (ICA) of the EEG signals, and fuzzy neural networks (FNNs) called  ICA-FNN. 

\begin{figure}
    \centering
    \tikzstyle{block} = [rectangle, draw,  minimum size= 3cm]
\tikzstyle{dotted_block} = [draw=black!30!white, line width=1pt, dash pattern=on 1pt off 4pt on 6pt off 4pt,
            inner ysep=6mm,inner xsep=4mm, rectangle, rounded corners ]
\tikzstyle{myarrows} = [ultra thick,->]
%-------------------------------------------------------------------
\pgfmathdeclarefunction{gauss}{2}{%
  \pgfmathparse{1/(#2*sqrt(2*pi))*exp(-((x-#1)^2)/(2*#2^2))}%
}
%-------------------------------------------------------------------
\begin{tikzpicture}[node distance = 1.5cm]
%------------------------------------------------
\hspace{-.5cm}
%---------- L1
\node(L1_C1)[circle, draw,  minimum size =.1cm,label = below:$x_1$, fill = green!30]{};
\node(L1_C2)[circle, draw, right of = L1_C1, minimum size =.1cm, yshift = 0cm,label = below:$x_2$, fill = green!30]{};
\node(dots1)[rectangle, draw=none, right of = L1_C2, minimum size =1cm, yshift = 0cm, rotate = 0, font = \Huge]{...};
\node(L1_C3)[circle, draw, right of = dots1, minimum size =.1cm, yshift = 0cm, label = below:$x_n$, fill = green!30]{};
%---------- L2
\foreach \i in {1,2,3}{
\node(L2_C\i_2)[circle, draw, above of = L1_C\i, minimum size =.1cm, node distance = .8cm,xshift = 0cm, yshift = 0cm, fill = blue!60]{};
\node(L2_C\i_1)[circle, draw, above of = L1_C\i, minimum size =.1cm,node distance = .8cm, xshift = -.4cm, yshift = 0cm, fill = blue!30]{};
\node(L2_C\i_3)[circle, draw, above of = L1_C\i, minimum size =.1cm,node distance = .8cm, xshift = .4cm, yshift = 0cm, fill = blue!90]{};
}
\node(dots2)[rectangle, draw=none, above of = dots1, minimum size =1cm, yshift = 0cm, rotate = 0,node distance = .8cm, font = \Huge]{...};
%---------- L3
\foreach \i in {1,2,3}{
\ifnum\i=3
\node(R_\i)[circle, draw, above of = L2_C\i_2, minimum size =.1cm, node distance = 1.5cm,xshift = 0cm, yshift = 0cm, fill = yellow!30]{\scriptsize $R_n$};
\else
\node(R_\i)[circle, draw, above of = L2_C\i_2, minimum size =.1cm, node distance = 1.5cm,xshift = 0cm, yshift = 0cm, fill = yellow!30]{\scriptsize $R_\i$};
\fi
}
\node(dots3)[rectangle, draw=none, above of = dots2, minimum size =1cm, yshift = 0cm, rotate = 0,node distance = 1.5cm, font = \Huge]{...};
%---------- L4
\foreach \i in {1,2,3}{
\node(L4_C\i_2)[circle, draw, above of = R_\i, minimum size =.1cm,xshift = 0.2cm, yshift = 0cm, node distance = 2cm, fill = red!30]{};
\node(L4_C\i_1)[circle, draw, above of = R_\i, minimum size =.1cm, xshift = -.2cm, yshift = 0cm, node distance = 2cm, fill=gray!30]{};
\node(L4_a\i)[ draw=none, left of = L4_C\i_1, minimum size =.01cm, yshift = -.5cm, node distance = .5cm]{\tiny \(\mathbf{x}\)};
}
\node(dots4)[rectangle, draw=none, above of = dots3, minimum size =1cm, yshift = 0cm, rotate = 0,node distance = 2cm, font = \Huge]{...};
%---------- L5
\node(L5_C1)[circle, draw, above of = R_2, minimum size =.1cm, yshift = 0cm, node distance = 3cm, fill = brown, xshift=1.5cm]{};
%-----------------------------------------------
%path 
% L1 to L2
%-----------------------------------------------
\foreach \i in {1,2,3}{
    \foreach \j in {1,2,3}{ 
        \draw[->](L1_C\i)--(L2_C\i_\j);};
};
%-----------------------------------------------
%L2 to L3
%-----------------------------------------------
\foreach \i in {1,2,3}{
    \foreach \j in {1,2,3}{ 
        \draw[->](L2_C\i_\j)-- (R_\j);};
};
%-----------------------------------------------
%within L3
%-----------------------------------------------
\foreach \i in{1,2,3}{
\foreach \j in {1,2,3}{
    \ifnum\i=\j
    \else
        \draw[->, dotted, bend left](R_\i) to (R_\j);
    \fi
};
};
%-----------------------------------------------
%L3 to L4
%-----------------------------------------------
\foreach \i in {1,2,3}{
    \foreach \j in {1,2}{ 
        \draw[->] (R_\i) -- (L4_C\i_\j);};
        %\draw[->] ++(-.1cm, -.1cm) |- (L4_C\i_\j);};
};
%-----------------------------------------------
% L4 to L5
%-----------------------------------------------
\foreach \i in {1,2,3}{
    \foreach \j in {1,2}{ 
        \draw[->] (L4_C\i_\j)-- (L5_C1);
            \ifnum \j=1
                \ifnum \i=3        
                    \draw[->] (L4_a\i)-- node [yshift=0.1cm, xshift =-.2cm]{\tiny \(a_n\)}(L4_C\i_\j);
                \else
                    \draw[->] (L4_a\i)-- node [yshift=0.1cm, xshift =-.2cm]{\tiny \(a_\i\)}(L4_C\i_\j);
                \fi
        \else
        \fi
        };
};
\draw[->] (L5_C1)-- node[yshift=.8cm]{$y_1$}++(0,1);
% layer names
\node(L1)[draw= none, left of = L1_C1, xshift = -.5cm]{Layer 1};
\node(L2)[draw= none, left of = L2_C1_1, xshift = -.1cm]{Layer 2};
\node(L3)[draw= none, left of = R_1, xshift = -.6cm]{Layer 3};
\node(L4)[draw= none, left of = L4_C1_1, xshift = -.4cm]{Layer 4};
\node(L5)[draw= none, left of = L5_C1, xshift = -3.6cm]{Layer 5};
\end{tikzpicture}
%-------------------------------------------------------------------
    \caption{The architecture of ICA-FNN by Lin \emph{et al.} \cite{CTLin_2006_ICAFNN} consists of five layers. The second layer transforms the inputs \(x_1\) to \(x_n\), obtained from the input layer 1, to latent variables computed using ICA on the input data. Each node in layer 3 is a rule (\(R_1\) to \(R_n\)), and calculates the antecedent match by computing the firing strength of each rule. Layer 4 is the consequent layer with defuzzification performed in layer 5 where a crisp output is produced.}
    \label{fig:ICAFNN}
\end{figure}
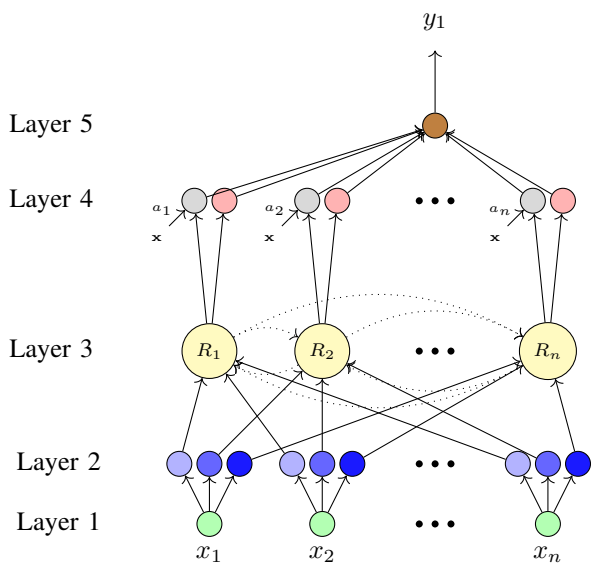

The significance of trying to decode the cognition state of alertness of an individual, based on the correlation between the information obtained from their brain signals, i.e., power spectra of ICA components of EEG signals, and the individual's driving performance, i.e., the difference between the centre of the vehicle and the cruising lane, is critical in alerting the driver before a potential car accident happens. This is of relevance to shed light on the brain development processes, as ICA-FNN can potentially be applied on infants' brain data to decode their cognitive states as defined by the (un)successful execution of the task at hand. 

The ICA-FNN architecture is defined over five layers, as shown in Fig. \ref{fig:ICAFNN}. Taken together, the fuzzy inference system of ICA-FNN takes the following form as shown in (\ref{eq:ICAFNN}):

\begin{align} \label{eq:ICAFNN}
    \begin{split}
    \text{Rule}: \; & \text{IF} \; antecedents \;\ \text{THEN} \ consequent(s)\\
    \text{Rule} \; i:\;& \text{IF} \; x_1 \; \text{is} \; A_1^i \; ... and \; x_j \; \text{is} \; A_j^i... and \; x_n \; \text{is} \; A_n^i  \\
                    & \text{THEN} \;y_i\; \text{is} \;m_{0i} \;+ \;a_{\textcolor{blue}{1i}}x_1 + \;... + \;a_{ji}x_j \;+ ... \;+ \;a_{ni}x_n
    \end{split}
\end{align}

where \(i\) is the rule number and \([x_1, ..., x_j, ...,x_n]\) are inputs, with conceptual labels defining the inputs as \([A_1^i, ..., A_j^i, ...,A_n^i]\) correspondingly becoming the antecedent part of the rule \(i\). The centre of a symmetric function is \(m_{0i}\), \(y_i\) is the consequent set, \(a_{ji}\) is a consequent parameter for the \(jth\) antecedent of the \(ith\) rule.

The antecedent part of the rules are latent variables obtained from ICA analysis of the EEG signals. Although the inference mechanism is expressible in terms of the rules acquired, see eq. (\ref{eq:ICAFNN}), the inputs are abstract features derived from EEG signals. In this regard, the rules can not shed light on the underlying cortical networks formed owing to the non-explainability of the abstract antecedents of the rules. 

In addition to the limited explainability of the rules obtained, the rules are also different across the subjects. This is because of the adaptive feature selection mechanism based on ICA; hence, the rules obtained can not be generalised across the subjects. % The rules for each subject can not be taken together to shed light on the underlying cognitive states of alertness across subjects. 
Moreover, each subject would also need to undergo the training phase separately to tune the hyperparameters using the backpropagation algorithm, to learn the cognitive states of each individual. In this way, the inter-subject variabilities of cognition states are largely accounted for and yield improved performance for the recognition of drowsiness/alertness. They reported a remarkable average accuracy of 98.2 \(\pm\) 1.0 \% over five subjects. 

%------------------------------------------------
%\subsection{XAI model of RepL using ICA-FNNs}
%-------------------------------------------
% The usage of FNNs % for recognition of cognition state of alertness
% in ICA-FNN model by Lin \emph{et al.} \cite{CTLin_2006_ICAFNN} allowed for the incorporation of a rule-based inference mechanism. The advantage of a rule-based inference mechanism is based on the easily interpretable patterns that the inference mechanism has learnt to shed light on the underlying brain activity patterns. However, in the work by Lin \emph{et al.} \cite{CTLin_2006_ICAFNN}, the rules are learnt and formed on the features obtained from the ICA of EEG signals. Since the input features are abstract; the antecedents of the rules are not explainable in terms of brain activity patterns rendering the ICA-FNN model to be only partially explainable. 

%The inference mechanism linked the abstract features in rules learnt separately for each person.
The rules are also learnt on-line, i.e., during the training phase for each subject. By learning the rules on-line the inter-subject variabilities are accounted for. This also helps for better recognition of drowsiness since ICA-FNN would remember the inference mechanism is learnt for each subject to better decode that particular subject's level of drowsiness. Also, for a particular subject 2, they reported the acquired correlation between the abstract features learnt and the state of drowsiness of subject 2 as 0.93 and 0.88. Their investigations also concluded that drowsiness related regions are generally found to be in parietal and occipital lobes. 

%The strength of the model is in the rules obtained which can explain how abstract features obtained from different inputs are interconnecting (forming antecedent part of the rules) with each other that is leading to the cognitive state of alertness or drowsiness.

For the application of ICA-FNN in DCN studies, the on-line learning of the hyperparameters specific to each subject would need to be modified, based on the task at hand, since infants will not be able to provide feedback about their cognition state. In addition, to address the inter- and intra-subject variabilities,  type-2 fuzzy frameworks can be utilised as explored in these works \cite{Saha_2017_CognitiveFailure, Ghosh_2018_MotorLearningTask}. Similarly, EFCMs estimated values of EC can be mapped generally to the specialisation and neural reuse of the DCN processes, however not much insight can be gained about the activation(s) of the individual cortical regions. This is mainly because of how EFCMs seek to find the optimal values of the EC, by trying to minimise the error between the estimated and the actual values of the fNIRS signals with the help of GA. Hence, not much could be inferred about which part of the cortex is, for example, more active from the optimised EC values.

In the next section, we review some of the AI methods as applied to DCN studies. 
%----------------------------------------------------------
\subsection{AI in Developmental Cognitive Neuroscience}
%----------------------------------------------------------
The de-facto standard for analysis of DCN studies is univariate analysis based on simple statistical tests, where the cortical regions most active in response to the presented stimulus is recognised, i.e., it is an activation based analysis. There is also a tendency of translating models used in adult research to DCN; however, this entails making some assumptions. In contrast, very few DCN studies have focused on decoding the multivariate patterns in brain activity of infants in response to the presented stimuli (such as \cite{Emberson_2017} which is a correlation based MVPA). In fact, there is an evident scarcity for undertaking AI methods in DCN research.

In this subsection, we review the non-explainable and explainable AI methods as applied to DCN studies for conducting MVPA. 

%----------------------------------------------------------
\subsubsection{Non-Explainable AI Method}
\emph{(a) MVPA with EEG using SVM}\\
%----------------------------------------------------------
In the study by Bayet \emph{et al.} \cite{Bayet_2020_TemporalMVPA}, time resolved EEG based MVPA is conducted using a linear SVM. Infants aged 12 to 15 months participated in the study. The aim of the study was to investigate whether neural representations in the adult brain are different from the developing brain for the processing of visual stimuli (animals vs human body parts). The group-wise classification results of the SVM based MVPA was able to successfully decode between infants' and adults' brain activation patterns in response to the presented stimuli. However, infant multivariate representations didn’t linearly separate for animal and body images.

The study was able to establish that neural representation for visual information processing, of animals vs body parts differ significantly between infants and adults. These findings were significant by suggesting that the cortical networks undergo the processes of localisation and specialisation to process the presented visual stimulus information. However, the study could not shed light on what cortical networks were activated for adults, and likewise, for infants, that could explain the underlying brain mechanism correspondingly. This is mainly because of the non-explainable inference mechanism of SVM as discussed previously in  \ref{sec:FC_EEG_SVM}.

%The next subsection presents an alternative MVPA method that solves the drawbacks or poor desciphrability and semantic meaning of the multivariate patterns.  The internal SVM mechanism is replaced by a concept based compute with words (CWW) approach.\\

\noindent
%----------------------------------------------------------
\emph{(b) MVPA with fNIRS using correlation}\\
%----------------------------------------------------------
In contrast to EEG's MVPA analysis (which is temporally driven), an fNIRS based MVPA is aimed at spatial investigations into the cortical regions' activations encoded in the MVM. A hypothetical construction of a MVM using six fNIRS signals reading from six different cortical regions is depicted in Fig. \ref{fig:xMVPA} (a) - (b). The work by Emberson et al. \cite{Emberson_2017} decoded the brain responses in 19 six-months-old infants' fNIRS signals in response to auditory and visual stimuli. They decoded the signals by undertaking a MVPA driven by correlation and reported an average classification accuracy of 66.67\% for trial-level decoding.

The significance of their work lies in usage of MVPA that improved the decoding sensitivity in comparison to their previous work that used univariate methods \cite{Emberson_2015}. A feature significance analysis was also undertaken to determine which features (fNIRS channels) are most significant for recognising the fNIRS signals in response to visual and auditory stimuli. Their results indicated channel 1 (occipital cortex), channel 3 (occipital cortex), and channel 8 (prefrontal cortex) to be the most critical channels for decoding between visual and auditory stimuli.

The identification of fNIRS channels and their corresponding anatomical locations sheds light on the localised activation of the cortex as delineated by the IS framework. In addition, the improved sensitivity of MVPA on account of analysing more than one variable (fNIRS channels' activity) rather than univariate analysis further corroborates that cortical networks (interaction between multiple cortical regions) are formed for the processing of perceptual stimuli. In this sense, the correlation based MVPA is able to implicitly imply the formation of cortical networks. However, what exactly entails the cortical networks is unknown because the presence and type of interaction between the fNIRS channels is unrevealed by the correlation based MVPA.

Motivated from the success of the correlation based MVPA by Emberson \cite{Emberson_2017} and to overcome its limitation of partial explainability, we designed an explainable MVPA (xMVPA) which is reviewed next. 
%----------------------------------------------------------
\subsubsection{XAI Method: MVPA with fNIRS using eXplainable MVPA (xMVPA) }
%----------------------------------------------------------
%------------------------------------------------
In order to retain the brain activity patterns in MVM during the learning of the classification mechanism of a given ML algorithm to drive MVPA, a previous work from our group \cite{AndreuPerez_2021_xMVPA} has explored using Fuzzy Logic to power MVPA, called eXplainable MVPA (xMVPA). The Fuzzy Logic System (FLS) is unique in its ability to compute with words (CWW) as well as account for the uncertainty in the input data by assigning a membership grade \(\mu\) in the range \([0,1]\) to each input value \(x\). 

In the work \cite{AndreuPerez_2021_xMVPA} an interval type-2 fuzzy logic system (IT2-FLS) \cite{Hagras_2018_xAI} is used for powering the MVPA to analyse the brain activity patterns of six-month-old infants in response to sensory inputs. The MVM, constructed from fNIRS channels of interest on the occipital (associated with visual processing), temporal (associated with auditory processing), and PFC (associated with thinking and planning), is first converted into a conceptual linguistic label (CoL) MVM based on the definition of the membership functions (MFs) of the CoLs, as illustrated in Fig. \ref{fig:xMVPA} (c) - (d). A mathematical definition of IT2 membership functions is given in eq. (\ref{eq:IT2FS}).

\begin{align} \label{eq:IT2FS}
\begin{split}
   \tilde{A}= \{(x,\mu,1)| &\forall x \in X, \\ &\forall \mu \in [\mu_{\tilde{A}}(x),\mu_{ \tilde{A}}(x)] \subseteq [0,1]\}
\end{split}
\end{align}

where \(\mu_{\tilde{A}}\) represent the MF of interval type-2 fuzzy set \(\tilde{A}\) defined over input \(x\).

%trim = L B R T
%----------------------------------------------------------
%----------------------------------------------------------
\tikzstyle{dotted_block} = [draw=black!30!white, line width=1pt, dash pattern=on 1pt off 4pt on 6pt off 4pt, inner ysep=2mm,inner xsep=4mm, rectangle, rounded corners ]
\tikzstyle{myarrows} = [ultra thick,->]
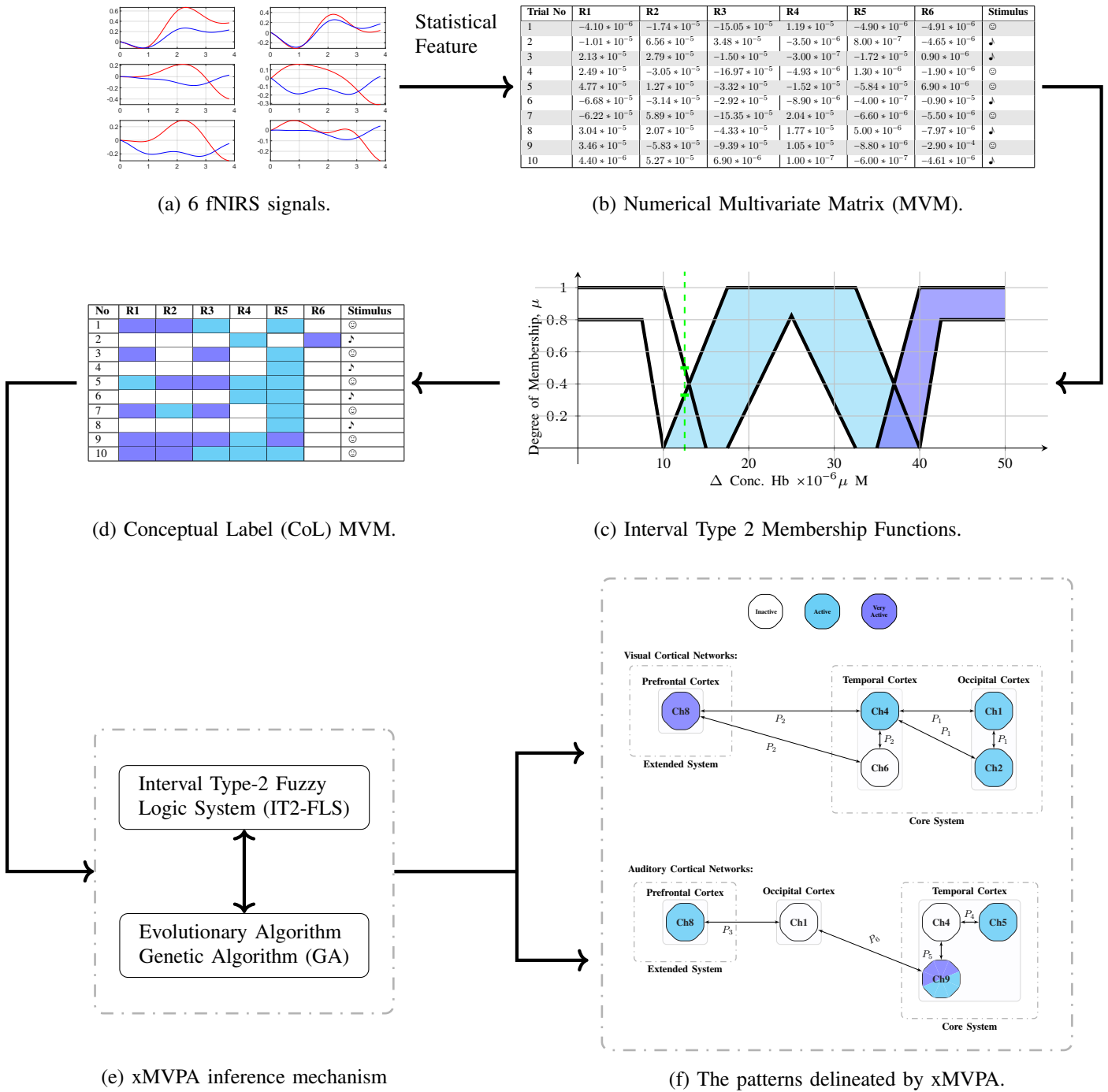
\begin{figure*}[!htbp]
\begin{tikzpicture}[node distance = 6cm]
\hspace{0cm}
%------------------------------------------------
\node(fsig)[draw= none]{\includegraphics[scale = 0.39, trim={4cm, 12.5cm, 4.2cm, 9.9cm}, clip=true]{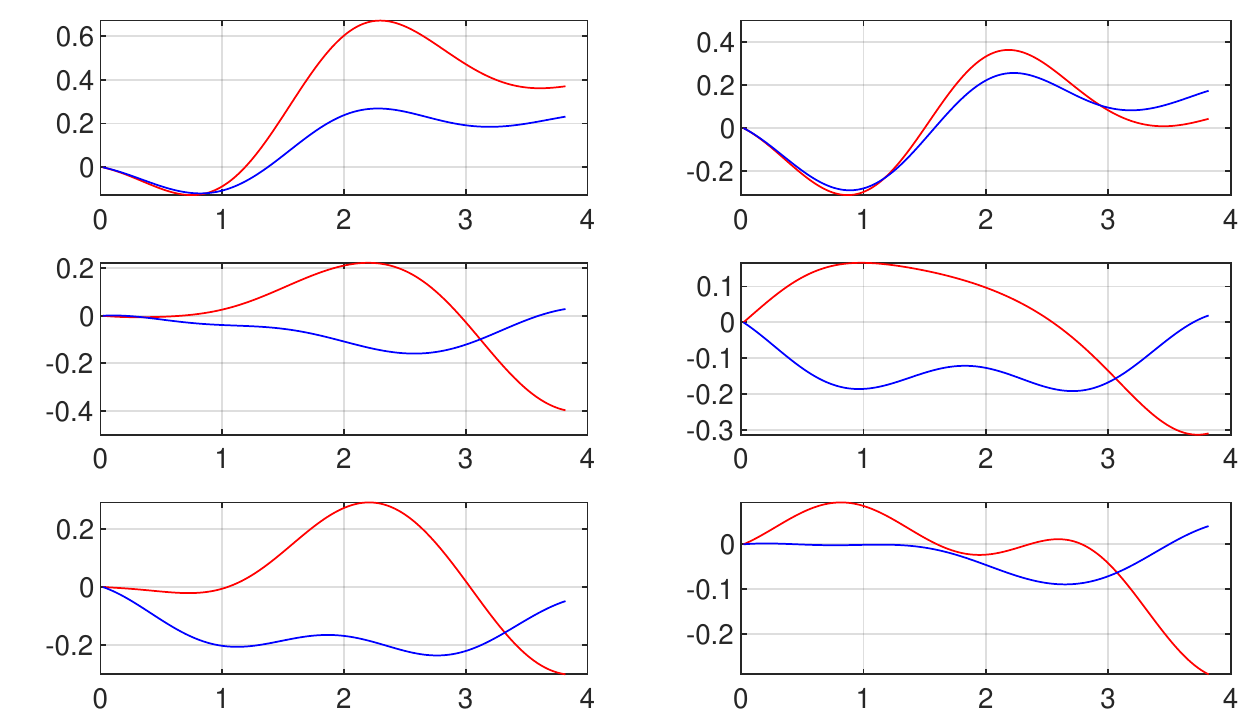}};
\node (fsig_caption) [draw = none, below of = fsig, node distance = 2cm]{(a) 6 fNIRS signals.};
%----------------------------------------------------------
\node(N_MVM)[draw= none, right of = fsig, xshift = 3cm] {% \begin{table}[!tbp]
% \centering
% \caption{An illustrative multivariate matrix (MVM) of brain activity from 5 different regions (R) for a total of 10 trials. The rows are trials, whereas the columns are the different regions from where brain activity is measured: channels (for fNIRS) or electrodes (for EEG). For fNIRS the cortical brain activity is recorded by a change in Hb concentration in the blood measured in \(\mu\)Mol and for EEG the recorded electrical brain activity is measured in \(\mu\)V.}
\scalebox{0.5}{\begin{tabular}{|l|l|l|l|l|l|l|l|} 
\hline
\textbf{Trial No} &\textbf{R1} 		&\textbf{R2}	& \textbf{R3} 		&\textbf{R4} 		&\textbf{R5} 		& \textbf{R6} & \textbf{Stimulus} 		\\ [0.5ex]\hline
\rowcolor{gray!20}1				&$-4.10*10^{-6}$	&$-1.74*10^{-5}$	&$-15.05*10^{-5}$	&$1.19*10^{-5}$	&$-4.90*10^{-6}$ &$-4.91*10^{-6}$	&  \smiley{}			\\ [0.5ex]
2				&$-1.01*10^{-5}$	&$6.56*10^{-5}$	&$3.48*10^{-5}	$	&$-3.50*10^{-6}$	&$8.00*10^{-7}$	&$-4.65*10^{-6}$	&  \textmusicalnote 	\\ [0.5ex]
\rowcolor{gray!20}3				&$2.13*10^{-5}$	&$2.79*10^{-5}$	&$-1.50*10^{-5}$	&$-3.00*10^{-7}$	&$-1.72*10^{-5}$ &$0.90*10^{-6}$		& \textmusicalnote 	\\ [0.5ex]
4				&$2.49*10^{-5}$	&$-3.05*10^{-5}$	&$-16.97*10^{-5}$	&$-4.93*10^{-6}$	&$1.30*10^{-6}$ &$-1.90*10^{-6}$			& \smiley{}			\\ [0.5ex] 
\rowcolor{gray!20}5				&$4.77*10^{-5}$	&$1.27*10^{-5}$	&$-3.32*10^{-5}$	&$-1.52*10^{-5}$	&$-5.84*10^{-5}$ &$6.90*10^{-6}$		&\smiley{}			\\ [0.5ex] 
6				&$-6.68*10^{-5}$	&$-3.14*10^{-5}$	&$-2.92*10^{-5}$	&$-8.90*10^{-6}$	&$-4.00*10^{-7}$ &$-0.90*10^{-5}$		& \textmusicalnote 	\\  [0.5ex] 
\rowcolor{gray!20}7				&$-6.22*10^{-5}$	&$5.89*10^{-5}$	&$-15.35*10^{-5}$	&$2.04*10^{-5}$	&$-6.60*10^{-6}$	&$-5.50*10^{-6}$	&\smiley{}			\\ [0.5ex] 
8				&$3.04*10^{-5}$	&$2.07*10^{-5}$	&$-4.33*10^{-5}$	&$1.77*10^{-5}$	&$5.00*10^{-6}$ &$-7.97*10^{-6}$	& \textmusicalnote 	\\ [0.5ex] 
\rowcolor{gray!20} 9				&$3.46*10^{-5}$	&$-5.83*10^{-5}$	&$-9.39*10^{-5}$	&$1.05*10^{-5}$	&$-8.80*10^{-6}$&$-2.90*10^{-4}$	&\smiley{}			\\ [0.5ex] 
10				&$4.40*10^{-6}$	&$5.27*10^{-5}$	&$6.90*10^{-6}$	&$1.00*10^{-7}$	&$-6.00*10^{-7}$	&$-4.61*10^{-6}$	& \textmusicalnote	\\ \hline 
\end{tabular}}
% \label{tab:MVPA_numerical_matrix}
% \end{table}
};
\node (N_MVM_caption) [draw = none, below of = N_MVM, node distance = 2cm]{(b) Numerical Multivariate Matrix (MVM).};
%----------------------------------------------------------
\node(MF)[draw= none, below of = N_MVM, yshift = 1cm]{%----------------------------------------------------------
\definecolor{red}{rgb}{0.89, 0.0, 0.13}
\definecolor{yellow}{rgb}{1.0, 0.75, 0.0}
%%%%%%%%%%%%%%%%%%%%%%%%%%%%%%%%%%%%%%%%%%%%%%%%%
% \begin{figure}[!tb] 
% \centering
\begin{tikzpicture}[scale=.6]
\begin{axis}[
axis lines = center, 
width=16cm,
height=7cm,
enlargelimits,
grid=both,
x label style={at={(axis description cs:0.8,-0.06)},anchor=north},
y label style={at={(axis description cs:-0.01,.7)},rotate=90,anchor=south},
yticklabel style={fill=white, font =\scriptsize},
xticklabel style={fill=white, font = \scriptsize},
xtick = {0, 0.2, 0.4, 0.6, 0.8, 1},
xticklabels={0, 10, 20, 30, 40, 50},
xlabel style={text width=3cm},        
xlabel =  {\scriptsize \(\Delta\) Conc. Hb \(\times 10 ^ {-6} \mu\) M},
ylabel = {\scriptsize Degree of Membership, \(\mu\)},
clip=false,
axis on top,
name=ax,
legend columns=3,
legend style={at={(.73,1.1)},draw=none,font =\scriptsize,
                    % the /tikz/ prefix is necessary here...
                    % otherwise, it might end-up with `/pgfplots/column 2`
                    % which is not what we want. compare pgfmanual.pdf
            /tikz/column 2/.style={
                column sep=20pt,
            },
            /tikz/column 4/.style={
                column sep=20pt,
            },
        },
%legend cell align={top},
%legend style={ at={(.65,1.1)},draw=none, font =\scriptsize},
]
\addplot[name path=Ll, ultra thick, black] coordinates {(0,0.8)(.15,0.8)(.2,0)};
\addplot[name path=Lu, ultra thick, black] coordinates {(0,1)(.2,1)(.3,0)};
\addplot[name path=Ml, ultra thick, black] coordinates {(.35,0)(.5,0.825)(.65,0)}; 
\addplot[name path=Mu, ultra thick, black] coordinates {(.2,0)(.35,1)(.65,1)(.8,0)}; 
\addplot[name path=Hl, ultra thick, black] coordinates {(.8,0)(.85,0.8)(1,0.8)}; 
\addplot[name path=Hu, ultra thick, black] coordinates {(.70,0)(.8,1)(1,1)}; 
\addplot+[fill=white,opacity=0.5] fill between[of=Ll and Lu,soft clip={domain=0:.5}];
\addplot+[fill=cyan!50,opacity=0.5] fill between[of=Ml and Mu,soft clip={domain=.1:.8}];
\addplot+[fill=blue!50,opacity=0.7] fill between[of=Hl and Hu,soft clip={domain=.2:1}];
 \draw[ thick, dashed, green] ({axis cs:.25,0}|-{rel axis cs:0,0.07}) -- ({axis cs:.25,0}|-{rel axis cs:0.6,1});
 \draw[green, ultra thick] (axis cs:.24,0.5) -- (axis cs:.26,0.5);
 \draw[green, ultra thick] (axis cs:.24,.33) -- (axis cs:.26,.33);
 %\draw[green, ultra thick] (axis cs:.23,1) -- (axis cs:.27,1);
% \legend{\textcolor{black}{Inactive}, ,\textcolor{yellow}{Active},  ,
% \textcolor{red}{Very Active}}
\end{axis}
\end{tikzpicture}
% \caption{\small{An illustrative plot to exemplify how conceptual labels (CoLs), that can be used to describe thermal comfort based on the room temperature (T \textdegree C), are characterised with uncertainty handling in xMVPA. Thermal comfort can be expressed with the CoLs \emph{cold}, \emph{comfortable} and \emph{hot} with approximate degree of membership values, \(\mu\), obtained from \cite{ChengdongLi_2013}. As can be seen in the figure, the definition of CoLs is not necessarily mutually exclusive, i.e. a certain temperature can be represented using more than one CoL with varying degrees of membership. For example, the temperature of 12 \textdegree C has degree of membership, \(\mu\), in the range of (0, 0.5) for \emph{cold} and (0, 0.33) for \emph{comfortable}. The derived ambiguity in the degree of membership ensures that uncertainty in the numeric data (or neuroimaging reading from fNIRS) is well retained upon transformation into a CoL.}}
% \label{Fig:MemFun}
% \end{figure} 
}; 
\node (MF_caption) [draw = none, below of = MF, node distance = 2.5cm]{(c) Interval Type 2 Membership Functions.};
%----------------------------------------------------------
\node(CoL_MVM)[draw= none,  left of = MF, xshift = -3cm, yshift = 0cm]{% \begin{table}[!htbp]
\scalebox{.55}{
\begin{tabular}{|p{.5cm}|p{.7cm}| p{.7cm} | p{.7cm} | p{.7cm}| p{.7cm}|  p{.7cm} |  p{1.5cm} |}
\hline
\textbf{No}	&\textbf{R1}	 & \textbf{R2}	& \textbf{R3} 	& \textbf{R4} & \textbf{R5} & \textbf{R6}	&\textbf{Stimulus}\\ \hline
1			& \cellcolor{blue!50}	 	& \cellcolor{blue!50}		& \cellcolor{cyan!50}		& \cellcolor{white}	 & \cellcolor{cyan!50} &	& \smiley{} 		\\ \hline
2			&\cellcolor{white}		& \cellcolor{white}		& \cellcolor{white}		& \cellcolor{cyan!50}	&  & \cellcolor{blue!50}	&\eighthnote\\ \hline
3			& \cellcolor{blue!50} 		& \cellcolor{white}		& \cellcolor{blue!50}		& \cellcolor{white}	& \cellcolor{cyan!50} &	&\smiley		\\ \hline
4			&\cellcolor{white}		& \cellcolor{white}		& \cellcolor{white}		& \cellcolor{white}	& \cellcolor{cyan!50} &	&\eighthnote\\ \hline
5			& \cellcolor{cyan!50} 		& \cellcolor{blue!50}		& \cellcolor{blue!50}		& \cellcolor{cyan!50}	& \cellcolor{cyan!50} &	& \smiley{}	\\ \hline
6			&\cellcolor{white}		&\cellcolor{white}		& \cellcolor{white}		& \cellcolor{cyan!50}	& \cellcolor{cyan!50} &	&\eighthnote\\ \hline
7			&\cellcolor{blue!50}	 	& \cellcolor{cyan!50}		& \cellcolor{blue!50}		& \cellcolor{white}	& \cellcolor{cyan!50}	&		&\smiley\\ \hline
8			&\cellcolor{white}		& \cellcolor{white}		& \cellcolor{white}		& \cellcolor{white}	& \cellcolor{cyan!50}	& &\eighthnote\\ \hline
9			& \cellcolor{blue!50}	 	& \cellcolor{blue!50}		& \cellcolor{blue!50}	& \cellcolor{cyan!50}	& \cellcolor{blue!50} &		&\smiley\\ \hline
10			& \cellcolor{blue!50}	 	& \cellcolor{blue!50}		& \cellcolor{cyan!50}		& \cellcolor{cyan!50}	& \cellcolor{cyan!50}	&	&\smiley\\ \hline
\end{tabular}}
    % \caption{An illustrative multivariate pattern matrix of brain activity from 5 different regions (R) for a total of 10 trials. The rows are trials, whereas the columns are the different regions from where brain activity is measublue!50: channels (for fNIRS) or electrodes (for EEG). Here, the colours denote the level of activity in different brain regions in response to the presented stimuli For example, blue!50 representing high activity, cyan!50 low activity, and white no activity.}
    % \label{tab:MVPA_matrix}
% \end{table}
};
\node (CoL_caption) [draw = none, below of = CoL_MVM, node distance = 2.5cm]{\begin{tabular}{l}
     (d) Conceptual Label (CoL) MVM.
\end{tabular} };
%------------------------------------------------------
%----------------------------------------------------------
\node(xMVPA_mech1)[ rectangle, draw,  below of = CoL_MVM, yshift = -1cm, rounded corners]{ \begin{tabular}{l}
     Interval Type-2 Fuzzy \\
     Logic System (IT2-FLS)
\end{tabular}};
\node(xMVPA_mech2)[rectangle,  below of = xMVPA_mech1,  yshift = 3.5cm, draw, rounded corners]{\begin{tabular}{l}
     Evolutionary Algorithm\\
     Genetic Algorithm (GA)
\end{tabular}};
\node(xMVPA_mech)[dotted_block,inner ysep=6mm,  fit =(xMVPA_mech1)(xMVPA_mech2) ]{};
\node (xMVPA_caption) [draw = none, below of = xMVPA_mech, node distance = 3.5cm]{\begin{tabular}{l}
     (e) xMVPA inference mechanism  
\end{tabular} };
%----------------------------------------------------------
\node(V_patterns)[draw= none,  right of = xMVPA_mech, yshift = 2.7cm, scale=0.48, node distance = 10cm]{\tikzstyle{channel_active} = [shape= regular polygon, draw, fill=cyan!50, regular polygon sides=8,rounded corners,
    text width=.7cm, text centered, minimum width=.5cm]
\tikzstyle{channel_inactive} = [shape= regular polygon, draw, fill=none, regular polygon sides=8,rounded corners,
    text width=.7cm, text centered, minimum width=.5cm, opacity=1]
\tikzstyle{channel_veryactive} = [shape= regular polygon, draw, fill=blue!50, regular polygon sides=8,rounded corners,
    text width=.7cm, text centered, minimum width=.5cm]
\tikzstyle{block} = [rectangle, draw, fill=blue!5, 
    text width=3.2cm, text badly centered, rounded corners, minimum height=1.5cm]
\tikzstyle{dotted_block} = [draw=black!30!white, line width=1pt, dash pattern=on 1pt off 4pt on 6pt off 4pt, inner ysep=6mm,inner xsep=4mm, rectangle, rounded corners ]
%\begin{figure}
%    \centering
    % \subfloat[A cortical network proposed by xMVPA for processing \textbf{visual} stimulus in 6-month-old infants.]{%{\label{fig:Haxby_xMVPA}}
    \begin{tikzpicture}[node distance = 2cm, auto]
        % Place nodes
        \hspace{-.5cm}
        \node [channel_active] (Ch1) {\textbf{Ch1}};  
        \node [channel_active, below of = Ch1] (Ch2) {\textbf{Ch2}};  
        \node [block, fit = (Ch1)(Ch2),  opacity=.15, text opacity=1, label=above:\textbf{Occipital Cortex}] (Occi) {};
        %\node [block, right of = Occi, node distance = 5cm, fill=green!20,] (temp) {\textbf{Temporal Cortex} \(Ch_4\) active};
        \node [channel_active, left of = Ch1, node distance = 4cm] (Ch4) {\textbf{Ch4}};  
        \node [channel_inactive, below of = Ch4, opacity=1] (Ch6) {\textbf{Ch6}};  
        \node [block, fit = (Ch4)(Ch6),  opacity=0.1, text opacity=1, label=above:\textbf{Temporal Cortex}] (temp) {};
       
       %\node [block, right of = temp, node distance = 7cm, fill=red!20, ] (PFC) {\textbf{Prefrontal Cortex} \(Ch_8\) very active};
        \node [channel_veryactive, left of = Ch4, node distance = 7cm] (Ch8) {\textbf{Ch8}};  
        \node [block, fit = (Ch8),  opacity=.15, text opacity=1, label=above:\textbf{Prefrontal Cortex}] (PFC) {};
        \node [dotted_block, fit = (Occi)(temp), label=below:\textbf{Core System}, inner xsep=9mm ,inner ysep=8mm] (Core) {};
        \node [dotted_block, fit = (PFC), label=below:\textbf{Extended System}, inner xsep=10mm ,inner ysep=8mm] (Extended) {};
         \node [block, text opacity=1, above of = PFC,node distance = 0.95cm, draw=none,opacity=0, xshift = 0cm,fill=white, label=above:\textbf{Visual Cortical Networks:}] (VP) {};
        % Legend
        \node [channel_inactive, above of = Ch8, node distance = 3.5cm,  minimum size=1.1cm, scale = 0.7, xshift = 4.25cm, inner sep = 3pt, text width = 1cm,text badly centered] (L_iA) {\small \textbf{Inactive}};
        \node [channel_active, right of = L_iA, node distance = 2cm, text badly centered, minimum size=1.1cm, scale = 0.7, text width = 1cm] (L_A) {\small\textbf{Active}}; 
        \node [channel_veryactive, right of = L_A, node distance = 2cm, text badly centered, text width = 1cm, minimum size=1.1cm, scale = 0.7,  font=\small\linespread{1}\selectfont] (L_VA) { \textbf{Very\\Active}};
        % Paths
        \draw [latex'-latex', thick, >=stealth] (Ch1) -- node[]{\(P_1\)}(Ch2);
        \draw [latex'-latex', thick, >=stealth] (Ch1) -- node[]{\(P_1\)}(Ch4);
        \draw [latex'-latex', thick, >=stealth] (Ch4) -- node[]{\(P_1\)}(Ch2);
        \draw [latex'-latex', thick, >=stealth] (Ch4) -- node[]{\(P_2\)}(Ch6);
        \draw [latex'-latex', thick, >=stealth] (Ch6) -- node[]{\(P_2\)}(Ch8);
        \draw [latex'-latex', thick, >=stealth] (Ch4) -- node[]{\(P_2\)}(Ch8);
\end{tikzpicture}};

\node(A_patterns)[draw=none,  below of =V_patterns, yshift = 2cm, scale =0.48]{%octagon/.style=
%  {shape=regular polygon, regular polygon sides=8, draw, minimum width=.8in}
\tikzstyle{channel_active} = [shape= regular polygon, draw, fill=cyan!50, regular polygon sides=8,rounded corners,
    text width=.7cm, text centered, minimum width=.5cm]
\tikzstyle{channel_inactive} = [shape= regular polygon, draw, fill=none, regular polygon sides=8,rounded corners,
    text width=.7cm, text centered, minimum width=.5cm, opacity=1]
\tikzstyle{channel_veryactive} = [shape= regular polygon, draw, fill=blue!50, 
    regular polygon sides=8,rounded corners,
    text width=.7cm, text centered, minimum width=.5cm]
\tikzstyle{block} = [rectangle, draw, fill=blue!5, 
    text width=3.2cm, text badly centered, rounded corners, minimum height=1.5cm]
\tikzstyle{dotted_block} = [draw=black!30!white, line width=1pt, dash pattern=on 1pt off 4pt on 6pt off 4pt, inner ysep=6mm,inner xsep=4mm, rectangle, rounded corners ]
\begin{tikzpicture}[node distance = 2cm, auto]
        % Place nodes
        % --- Temporal Cortex
        \hspace{-.5cm}
        \node [channel_inactive] (Ch4) {\textbf{Ch4}};  
        \node [channel_active, right of = Ch4] (Ch5) {\textbf{Ch5}};
        %\node[name=Ch9,
        % shape=circle split,
        % text width=1cm, text centered,  minimum size=.8cm,
        % draw=gray!40,line width=.01mm,
        % circle split part fill={cadmiumred,amber}, below of = Ch4, rotate=90, node distance = 2.5cm, yshift=-1.2cm, shape border rotate=0] {};
         \node [channel_inactive, fill=none, below of = Ch4] {}; 
        \polygon[below of = Ch4,  minimum width=.5cm]{8}{.7}{0,1,2,3}{blue!50}
        \polygon[below of = Ch4,  minimum width=.5cm]{8}{.7}{4,5,6,7}{cyan!50}
        \node [draw =none, below of = Ch4, node distance =2cm] {\textbf{Ch9}};
        \node [channel_inactive, draw= none, below of = Ch4] (Ch9) {}; 
        % \node [channel_active, below of = Ch4] (Ch9A) {\textbf{Ch9}};
        % \node [channel_veryactive, left of = Ch9A] (Ch9VA) {\textbf{Ch9}};
        \node [block, fit = (Ch4)(Ch5)(Ch9),  opacity=.15, text opacity=1, label=above:\textbf{Temporal Cortex}] (temp) {};
        
        % --- Occipital Cortex
        \node [channel_inactive, left of = Ch4, node distance = 5cm] (Ch1) {\textbf{Ch1}};  
        \node [block, fit = (Ch1),  opacity=.15, text opacity=1, label=above:\textbf{Occipital Cortex}] (Occi) {};
        % --- PreFrontal Cortex
        \node [channel_active, left of = Ch1, node distance = 4cm] (Ch8) {\textbf{Ch8}}; 
        \node [block, fit = (Ch8),  opacity=.15, text opacity=1, label=above:\textbf{Prefrontal Cortex}] (PFC) {};
        \node [dotted_block, fit = (temp), label=below:\textbf{Core System}, inner xsep=6mm ,inner ysep=6mm] (Core) {};
        \node [dotted_block, fit = (PFC), label=below:\textbf{Extended System}, inner xsep=10mm ,inner ysep=6mm] (Core) {};
        \node [block, text opacity=1, above of = PFC, draw=none,opacity=0, xshift =0.2cm,fill=white, node distance =.8cm, label=above:\textbf{Auditory Cortical Networks:}] (AP) {};
        % Legend
        % \node [channel_inactive, below of = Ch1, node distance = 3.5cm, text badly centered, minimum size=1.1cm, scale = 0.8] (L_iA) {\scriptsize \textbf{Inactive}}; 
        % \node [channel_active, right of = L_iA, node distance = 2cm, text badly centered, minimum size=1.1cm, scale = 0.8] (L_A) {\scriptsize \textbf{Active}}; 
        % \node [channel_veryactive, right of = L_A, node distance = 2cm, text badly centered, text width = 1cm, minimum size=1.1cm, scale = 0.75] (L_VA) {\scriptsize \textbf{Very Active}};
        % Paths
        \draw [latex'-latex', thick, >=stealth] (Ch1) -- node[xshift = -.5cm]{\(P_3\)}(Ch8);
        \draw [latex'-latex', thick, >=stealth] (Ch4) -- node[]{\(P_4\)}(Ch5);
        \draw [latex'-latex', thick, >=stealth] (Ch4) -- node[xshift = -.8cm, yshift=-.15cm] {\(P_5\)}(Ch9);
        \draw [latex'-latex', thick, >=stealth] (Ch1) -- node[rotate=30] {\(P_6\)}(Ch9);
        % \draw [latex'-latex', thick, >=stealth] (Ch6) -- node[]{\(P_2\)}(Ch8);
        % \draw [latex'-latex', thick, >=stealth] (Ch4) -- node[]{\(P_2\)}(Ch8);
\end{tikzpicture}};
\node(patterns)[dotted_block,  fit =(A_patterns)(V_patterns),inner ysep=2mm,inner xsep=2mm  ]{};
\node (pattern_caption) [draw = none, below of = patterns, node distance = 4.5cm]{\begin{tabular}{l}
     (f) The patterns delineated by xMVPA.
\end{tabular} };
%------------------------------------------------------
% paths 
\draw[myarrows] (fsig) -- node[above, yshift=.3cm]{\begin{tabular}{l}
     Statistical\\
     Feature 
\end{tabular}}(N_MVM);
\draw[myarrows] (N_MVM.east)-- ++(1cm,0cm) |- (MF);
\draw[myarrows] (MF) -- (CoL_MVM);
\draw[myarrows] (CoL_MVM)-- ++(-4cm,0cm) |- (xMVPA_mech) ;
\draw[myarrows] (xMVPA_mech)-- ++(4.6cm,0cm) |- ([xshift=-.5cm, yshift = -.7cm]V_patterns.west) ;
\draw[<->, ultra thick] (xMVPA_mech1) -- (xMVPA_mech2);
\draw[myarrows] (xMVPA_mech)-- ++(4.6cm,0cm) |- ([xshift=-.5cm, yshift = -.2cm]A_patterns.west) ;
\end{tikzpicture}
\caption{(a) Six hypothetical fNIRS signals (b) An illustrative multivariate matrix (MVM) of brain activity from 6 different regions (R) for a total of 10 trials. The rows are trials, whereas the columns are the different regions from where brain activity is measured.(c) An illustrative plot to exemplify how conceptual labels (CoLs) can be used to represent brain activity. Brain regions can be expressed with the CoLs \emph{Inactive}, \emph{Active} and \emph{Very Active} with approximate degree of membership values, \(\mu\). The derived ambiguity in the degree of membership ensures that uncertainty in the numeric data (or neuroimaging reading from fNIRS) is well retained upon transformation into a conceptual label (CoL). (d) The corresponding CoL MVM. Here, the colours denote the level of activity in different brain regions in response to the presented stimuli. For example, violet representing high activity, cyan low activity, and white no activity. (e) The CoL MVM is given as an input to xMVPA which computes the classification accuracy of a given set of patterns on the given data by computing with words (CWW) using interval type-2 fuzzy logic system (IT2-FLS). The optimal set of patterns is found using an evolutionary algorithm (a genetic algorithm is used in the study). (f) A model for visual and auditory processing in six-month-old infants, based on the six patterns (\(P_1\) - \(P_6\)) revealed by the xMVPA inference mechanism \cite{AndreuPerez_2021_xMVPA}. The colour of the channel's (Ch) octagon is based on its activity level: Inactive (white), Active (cyan), and Very Active (violet).}
\label{fig:xMVPA}
\end{figure*}
%-------------------------------------------------------------
%-------------------------------------------------------------
%-------------------------------------------------------------

The CoL MVM is then fed to an evolutionary algorithm (such as GA) to find optimal patterns in the CoL MVM (train dataset) such that maximum classification accuracy can be obtained on the test dataset of the CoL MVM ( Fig. \ref{fig:xMVPA} (e)). The discerned patterns by the xMVPA are able to shed light on the activations and interconnections of the activated regions in the cortex in response to the presented stimuli. A general nomenclature of a pattern discerned by xMVPA is given in eq. (\ref{eq:pattern}).

\begin{align} \label{eq:pattern}
    \begin{split}
        Pattern \; P_m:\; & IF \; x_1 \; is \; A_{q_1} \;AND \;. . . AND \; x_n \; is \;A_{q_n}\\
        & THEN \; class \; is \; C_j \; with \ PW_m 
    \end{split}
\end{align}

where \(x_i\) are the crisp brain activity values for the variable \(i\) (fNIRS channel), \(A_i\) is the antecedent set for the \(i_{th}\) variable with a total of n variables and \(j \in J\) number of output classes. The class predicted with the pattern is \(C_j\) with the pattern weight given by \(PW_m\). The pattern weight, \(PW_m\), is a measure of the relevance of a given pattern as deemed by xMVPA based on its confidence, c, and support, s, values. The confidence can be viewed as the conditional probability that, given the antecedent(s) of \(P_m\), how likely the predicted class will be \(C_j\), whereas support measures the coverage (based on the number of matching data instances) of training dataset by the pattern \(P_m\). The \(\overline{upper}\) and \(\underline{lower}\; PW_m\) are defined as outlined in eq. (\ref{eq:PW}). 

\begin{align} \label{eq:PW}
\begin{split}
    \overline{PW_m} =& \overline{c_m} *\overline{s_m} \\
    \underline{PW_m} =& \underline{c_m} *\underline{s_m}
\end{split}
\end{align}

where \(c_m\) and \(s_m\) are confidence and support for the pattern \(P_m\) which can be calculated as outlined in \cite{Antonelli_2017_MultiObjective}.

The measurement of PW for each pattern enables a comparison to evaluate the efficacy of each learnt pattern for the given classification task. In the study by Andreu-Perez \emph{et al.} \cite{AndreuPerez_2021_xMVPA} a total of six patterns were identified for the processing of sensory stimuli that were able to recognise brain activity patterns of six-month-old infants, for the visual and auditory stimuli, encoded in CoL MVM that obtained a classification accuracy of 67.69\%. In \cite{AndreuPerez_2021_xMVPA}, the patterns as well as the MF definitions (start, height and endpoints) were learnt using GA with a total number of hyperparameters to be learnt equal to 300.

The xMVPA delineated patterns for visual processing that shed light on an occipital-temporal network as a core system that undertakes the primary processing of facial features, and of the PFC as an extended system that processes the emotion associated with the visual stimulus, as shown in Fig. \ref{fig:xMVPA} (f). The proposed model for auditory information processing consists of the temporal cortex as a core system for processing non-speech auditory stimuli, and of the PFC as an extended system that processes the emotion associated with the auditory stimulus, also shown in Fig. \ref{fig:xMVPA} (f). %
\section{Discussion} \label{sec:discussion}
%-------------------------------------------------------------
%\cellcolor{Gray}
\definecolor{Gray}{gray}{0.9}
% old table
\begin{table*}[!t]
\centering
\caption{A comparison of different artificial intelligence (AI) methods used on adult(A) and infant(I) neuroimaging data for Connectivity Analysis (CA), Representation Learning (RepL) and Multivariate Pattern Analysis (MVPA) with fNIRS and EEG data. The strengths (S) and limitations (L) of the AI methods are summarised. For CA, separate AI methods most commonly used with functional connectivity (FC), and effective connectivity (EC) are reported. The full name of the algorithms reviewed are: Support Vector Machine (SVM), Random Forest (RF), 1Dimensional Convolutional Neural Network with Long Short Term Memory (1DCNN-LSTM), Ridge Regression (RR), Integrated Functional Manifold (IFM), EEGNet based Convolutional Neural Networks (EEGNet), Symbol-Concept Association Network (SCAN), Convolutional Neural Networks (CNN), Effective Fuzzy Cognitive Maps (EFCM), Induced Type-2 Fuzzy Deep Brain Learning Network (IT2FDBN), Independent Component Analysis based Fuzzy Neural Network (ICA-FNN) and eXplainable MVPA (xMVPA).
%For each of these algorithms accuracy (ACC.) or the mean square error (MSE) depending on whether it was a classification or regression problem is also tabulated.
}
\label{tab:comparison_methods}
\renewcommand{\arraystretch}{1.5}
\scalebox{1}{
\begin{tabular}{l l l l c}
    & \textbf{Analysis Paradigm} & \textbf{Artificial Intelligence (AI) Methods} & \hspace{.5cm}\textbf{STRENGTHS (S) and LIMITATIONS (L)}  & \textbf{Population}\\     \hline
    %-------------------------------------------------------
    \multirow{14}{*}{\rotatebox[origin=c]{90}{Non-Explainable AI (non-XAI) Methods}} & \multirow{4}{*}{CA with EEG} & FC: SVM \cite{Moezzi_2019_EEG_SVM_FC}  \cite{ Klados_2020_PersonalityProfiles_EEG_FC_SVM}, RF: \cite{Kamarajan_2020_EEGwithRF}&\begin{tabular}{l}
        \hspace{-.2cm}\textbf{S:} Robust regression mechanism.\\
        \hspace{-.2cm}\textbf{L:} Limited spatial localisation. 
    \end{tabular} & \textbf{A}\\
%-------------------------------------------------------
    & &\cellcolor{Gray} EC: 1DCNN-LSTM  \cite{Saeedi_2020_EC_EEG_CNN} & \cellcolor{Gray}\begin{tabular}{l}
        \hspace{-.2cm}\textbf{S:} Automated feature learning.\\
        \hspace{-.2cm}\textbf{L:} Dependency on large datasets. 
    \end{tabular}  & \cellcolor{Gray} \textbf{A} \\ \cline{2-5}
    %-------------------------------------------------------
    %-------------------------------------------------------
    &  \multirow{4}{*}{CA with fNIRS} & FC: RR \cite{Duan_2020_FC_fNIRS_CorticalNetworks}& \begin{tabular}{l}
     \hspace{-0.2cm}\textbf{S:} Simple analytical model.\\
     \hspace{-0.2cm}\textbf{L:} Results are dependent on regressors (\(\beta\)).
     \end{tabular} & \textbf{A} \\
    %-------------------------------------------------------
    &  & \cellcolor{Gray}FC: IFM \cite{AvilaSansores_2020_IFM} &\cellcolor{Gray}\begin{tabular}{l}
     \hspace{-.2cm}\textbf{S:} Groupwise exploration is possible.\\
     \hspace{-.2cm}\textbf{L:} Manifold assumption.\\
     \end{tabular} & \cellcolor{Gray} \textbf{A}\\\cline{2-5}
    %-------------------------------------------------------
     & RepL with EEG & \begin{tabular}{l }
            \hspace{-.2cm}EEGNet (CNN) \cite{Lawhern_2018_EEGNet} \\ %\cite{Mengni_2018_EpilepticSeizure}
            \hspace{-.2cm}SCAN (CNN) \cite{Honke_2020_EEGRL} \\
            \end{tabular} &  \begin{tabular}{l}
            \hspace{-.2cm}\textbf{S:} Automated feature learning.\\
            \hspace{-.2cm}\textbf{L:} Dependency on large datasets.            \\
            \end{tabular}& \textbf{A} \\ \cline{2-5}
     %-------------------------------------------------------
     & RepL with fNIRS & \cellcolor{Gray} CNN  \cite{Saadati_2019_fNIRS_CNN}
    , \cite{Tanveer_2019_DrowsinessDetection_DL_fNIRS}, 
    \cite{Sasikala_2020_DCN_fNIRS_BCI}
    &\cellcolor{Gray}\begin{tabular}{l}
    \hspace{-.2cm}\textbf{S:} Automated feature learning.\\
    \hspace{-.2cm}\textbf{L:} Dependency on large datasets. 
    \end{tabular} &\cellcolor{Gray} \textbf{A} 
%https://arxiv.org/abs/2010.15274 -- Google
% they are trying to generate EEG data, that according to them it's more explainable, apparently they failed..., but here is their attempt 
  \\   \cline{2-5} 
    %-------------------------------------------------------
    %-------------------------------------------------------
    %-------------------------------------------------------
                                    & MVPA with EEG  & 
        SVM \cite{Cauchoix_2014_NeuralDynamicsOfFaceDetection} \cite{Bayet_2020_TemporalMVPA} %\\ 
        &\begin{tabular}{l}
            \hspace{-.2cm}\textbf{S:} Robust classification mechanism. \\
            \hspace{-.2cm}\textbf{L:}  Limited spatial localisation.  \\
            \end{tabular} & \textbf{A} + \textbf{I}\\  \cline{2-5} 
        %----------------------------------------
    %-------------------------------------------------------
     & MVPA with fNIRS  & 
        Correlation \cite{Emberson_2017} %\\ 
        &\begin{tabular}{l}
             \hspace{-.2cm}\textbf{S:} Better spatial localisation with infant level \\ and trial level decoding results.\\
            \hspace{-.2cm}\textbf{L:} Lack of insight into the interactions between \\ the important fNIRS channels. \\
            \end{tabular} & \textbf{I}\\   \hline
        %----------------------------------------
    %-------------------------------------------------------
     \multirow{5}{*}{\rotatebox[origin=c]{90}{\begin{tabular}{l}Explainable AI (XAI) Methods \end{tabular}}}     & CA with fNIRS & \cellcolor{Gray} EC: EFCM \cite{Kiani_2019_EFCM} &\cellcolor{Gray}\begin{tabular}{l}
     \hspace{-.2cm}\textbf{S:} Learnt parameters are EC. \\\hspace{-.2cm}\textbf{L:} Performance non-scalable.\end{tabular} & \cellcolor{Gray}\textbf{A}  \\ \cline{2-5}
    %-------------------------------------------------------
     &\multirow{4}{*}{FNNs %( Fuzzy Neural Networks) 
    with EEG} & IT2-FDBL \cite{Ghosh_2020_T2FDBL} &\begin{tabular}{l}
            \hspace{-.2cm}\textbf{S:} The empirical model proposed mimics short term \\ 
            memory of the brain. \\
            \hspace{-.2cm}\textbf{L:} Dependency on large datasets.  \\
            \end{tabular}&  \textbf{A} \\
       &      & \cellcolor{Gray} ICA-FNN \cite{CTLin_2006_ICAFNN} &\cellcolor{Gray} \begin{tabular}{l}
            \hspace{-.3cm}\textbf{S:} Automatic feature selection with uncertainty\\ handled with fuzzy models. \\
            \hspace{-.3cm}\textbf{L:} The obtained rules are not comparable per se \\
            because of online definition of linguistic labels.  \\
            \end{tabular} & \cellcolor{Gray}  \textbf{A} \\\cline{2-5}
%-------------------------------------------------------
% \multirow{6}{*}[16pt]{\rotatebox[origin=c]{90}{\begin{tabular}{r}
%          Infant Population 
%     \end{tabular}}}    
    % \multirow{1}{*}[10pt]{\rotatebox[origin=l]{90}{\begin{tabular}{c}
    %      non-XAI  \\
    %      Method 
    % \end{tabular}}}  & MVPA with EEG & SVM \cite{Bayet_2020_TemporalMVPA}&\begin{tabular}{l}
    %         \hspace{-.2cm}\textbf{S:} A comparison of developed and developing brain mechanisms. \\
    %         \hspace{-.2cm}\textbf{L:} The inference mechanism could not inform about the \\ underlying brain mechanisms. \\
    %         \end{tabular} \\  \cline{1-4}
%-------------------------------------------------------
    % \multirow{1}{*}[13pt]{\rotatebox[origin=l]{90}{\begin{tabular}{c}
    %      \hspace{2cm}  XAI  \\
    %      \hspace{2cm}  Method 
    % \end{tabular}
    % }}
    & MVPA with fNIRS &   xMVPA \cite{AndreuPerez_2021_xMVPA}&  \begin{tabular}{l}
            \hspace{-.3cm}\textbf{S:} Brain activity patterns defined by CWW.\\
            \hspace{-.3cm}\textbf{L:} Linguistic variables are determined a priori.\\
            \end{tabular}& \textbf{I}   \\  \hline
%-------------------------------------------------------
\end{tabular}
}
\end{table*}
%-------------------------------------------------------
%-------------------------------------------------------
%------------ TABLE with comparison --------------------
%-------------------- END ----------------------------
%-------------------------------------------------------
The implications for a greater insight into the developing brain mechanisms, both structural (physical) and functional (cognitive), are profound. According to a study in 2007, one in every four people is affected by a mental illness either directly or indirectly \cite{Webpage_NIH_MentalIllness}. In this regard, the study of a developing brain can inform us about \emph{typical and atypical brain developmental trajectories} which may in turn facilitate the early \emph{identification of and intervention for brain disorders} during childhood. However, at present, the DCN research has limited translation into shaping brain development because of typically small datasets (owing mostly to non-cooperating infant behaviour), the non-availability of \emph{a-priori} information about the underlying mechanisms in a developing brain, as well as a lack of explainability %(as defined below) 
of the AI techniques applied for the analysis of infant's neuroimaging data. 

% \begin{itemize}
%     \item Explainability
%     \begin{itemize} \label{list:exp}
%         \item What patterns of brain activity were observed in the neuroimaging data on whose basis is the inference mechanism learnt. Moreover, what insight can be obtained from the learnt inference mechanism about the underlying cortical brain networks.
%     \end{itemize}
% \end{itemize}

The neuroimaging modalities of fNIRS and EEG are considered to be the most `infant friendly', and have reached a pinnacle in their own right, where real time neuronal activity can be recorded with EEG, or can be localised to a corresponding anatomical location within 2cm using fNIRS. To benefit from their complimentary high resolutions, more recent studies have undertaken multimodal (fNIRS-EEG) brain activity analysis to gain greater insights into functional brain development. Likewise the improvement in AI methods, in particular, the revolution of ANNs into DNNs and the remarkable feature learning ability of CNNs %to conjure higher level features from low levels one
with hierarchical networks has led to breakthroughs in many challenging image classification, and speech recognition problems \cite{LeCun_2015_DeepLearning}. However, despite the advent of advanced neuroimaging technologies and the availability of sophisticated CNNs, the DCN research has not benefited as much from the aforementioned technological and computational advances in comparison to other complex fields (such as image classification). 

To this end, to bridge the gap between the information recorded by the neuroimaging technologies and the insights acquired from the analysis of the neuroimaging data, a review of the most prevalent AI techniques in different analysis paradigms is undertaken in the present work. In particular, the AI methods are investigated for their similarity with theoretical frameworks for DCN including Interactive specialisation (IS) \cite{Johnson_2001} and the neuroconstructivist approach \cite{KarmiloffSmith_2009} (summarised in section \ref{sec:IS}). The main processes in brain development include: 1) localisation 2) specialisation 3) parcellation and 4) neural reuse; and if an AI technique's learning mechanism can shed light on these DCN processes, it can then inform us about the underlying mechanisms of a developing brain in line with the aforementioned DCN processes.

In this regard, the inherent limitation of most AI methods to not be able to explain what was \emph{observed} in terms of brain activity patterns, during learning of their inference mechanism, renders them ill-suited to shed light on the DCN processes despite obtaining remarkable classification performance. A comparison of the strength and limitations of the AI methods, reviewed in this work, is summarised in Table \ref{tab:comparison_methods}. Indeed, the bottleneck is not the classification prowess of the reviewed AI methods owing to their advanced learning techniques to acquire abstract representations from the input data. The limitation of AI methods, as applied to DCN studies, is that without \emph{explainability} of the learnt inference mechanism, not much insight can be gained on the activated cortical regions for a given task. Of the AI methods reviewed, the only XAI method in DCN, to the best of the author's knowledge, is fuzzy logic based xMVPA \cite{AndreuPerez_2021_xMVPA}. 

The capability of fuzzy logic in  CWW (computing with words), and modelling uncertainty are particularly well-suited for neuroimaging data which is characterised by inter-subject variabilities. The xMVPA was able to discern patterns in the neural underpinnings of audio and visual processing in six-month-old infants. xMVPA is explainable since it identifies the patterns in the input data prototypical to the presented stimuli. The classification accuracy reflects the validity of the discerned patterns to represent the true brain activity patterns in correspondence to each stimulus. The learnt patterns are also explainable as they inform about the activations and interactions of the cortical areas. 

%interpret the results
The discerned patterns for the visual and auditory processing are illustrated in Fig. \ref{fig:xMVPA} (f). The cortical network formed for visual processing has a hierarchical structure with the processing of raw data processed in the occipital cortex, and the processed information is then passed to PFC where higher level processing is done. This pattern is widely observed in adult literature of visual processing \cite{Haxby_2000}, and verified the patterns found by xMVPA for processing of a visual stimulus in six-month-old infants. Based on the similarity of the cortical network formed for the visual stimulus with those of adults, the cortical network is identified as specialised. 

In contrast, the cortical network formed for auditory processing is hypothesised to be a non-specialised cortical network. The authors \cite{AndreuPerez_2021_xMVPA} suggested the formed network to be non-specialised based on the `inactive' activation status of a channel (Ch1) forming the link between the information pathway from the temporal cortex to PFC. This non-specialised network was hitherto unknown in DCN literature and was only discerned because of the explainable attributes of the xMVPA.

Likewise, xMVPA can shed light on interactions and activations for time resolved brain activity. To investigate the process of neural reuse, the patterns would need to be established for different time points. In this regard, an analysis of the interconnections that were present at a given time point and how these interconnections rewired to acquire a new cognitive or behavioural state at a later time point can be investigated in line with the neural reuse process of functional brain development. 

%-------------------------------------------------------------
\section{Conclusion} \label{sec:conclusion}
%-------------------------------------------------------------

Cognitive developmental delays and abnormalities are commonly associated with behavioural disorders that can become challenging conditions to treat in adulthood (\emph{such as attention deficit hyperactivity disorder, autism spectrum or bipolar disorders}) \cite{Feil_2016_EarlyIntervention}. The application of AI for cognitive neuroscience can extend novel ways of interrogating brain function by maximizing neuroimaging data. This is of particular interest for the study of developmental brains where the classical assumptions of brain function for adults cannot serve as guidance.

In this paper, the aim was to highlight the current gap in DCN research due to non-explainable AI methods. Since there is no insight obtained on the learnt classification mechanism on the basis of brain activity patterns, this critically limits the translation of DCN research to shape developing brain trajectories despite acquiring statistically significant classification results. To bridge the gap between DCN research and the translation of their insight(s), we suggest that future DCN research adopt a more explainable classification mechanism using XAI methods such as xMVPA \cite{AndreuPerez_2021_xMVPA}.

% The shift from univariate to multivariate data analysis has already The multivariate analyses of distributed patterns of brain responses is fundamental to investigation of which cortical areas are involved in the processing of presented information. In this regards, multivariate analysis can shed light on the localisation and specialisation of the brain regions as delineated by IS. 

%-------------------------------------------------------------
%\bibliographystyle{IEEEtran}
%\bibliography{MyReferences}
\renewcommand*{\bibfont}{\footnotesize}
\printbibliography
%-------------------------------------------------------------
%-------------------------------------------------------------
\end{document}